# Two-Factor Model of Soil Suction from Capillarity, Shrinkage, Adsorbed Film, and Intra-aggregate Structure

V.Y. Chertkov*

*Division of Environmental, Water, and Agricultural Engineering, Faculty of Civil and Environmental Engineering, Technion, Haifa 32000, Israel*

**Abstract**: The objective of this work is to derive the soil water retention from the soil structure without curve-fitting and only using the physical parameters found irrespective of an experimental retention curve. Two key points underlie the work: (i) the soil suction at drying coincides with that of the soil intra-aggregate matrix and contributive clay; and (ii) both the soil suction and volume shrinkage at drying depend on the same soil water content. In addition the two following results are used: (i) the available two-factor (capillarity and shrinkage) model of clay suction enables one to connect a clay suction and clay water content using the clay matrix structure; and (ii) the recent reference shrinkage curve model based on the concepts of intra-aggregate soil structure permits one to connect the soil water content at shrinkage with the water content of the contributive clay. With that the available two-factor model was essentially modified and, in particular, the effect of adsorbed water film was taken into account. The developed model includes the following input parameters: the solid density, relative volume of contributive-clay solids, relative volume of contributive clay in the oven-dried state, soil clay content, aggregate/intra-aggregate mass ratio, and specific volume of lacunar pores in the aggregates at maximum swelling. The validation of the model is based on available data of water retention and the above input parameters for six soils. A promising agreement between the predicted and observed water retention curves was found.

## 1. INTRODUCTION

Water retention is a key soil property. Methods of its measurement are known (e.g., [1,2]). However, its physical prediction, i.e., from a finite number of physical soil parameters that are obtained irrespective of soil water retention, is lacking. Different available models relating to swelling and non-swelling soils and originating from different empirical or physical considerations (e.g., [3-15]), are eventually reduced to curve-fitting to relevant experimental soil water retention data. Tuller et al. [7] and Or and Tuller [8] use parameters of a pore-size distribution in the fitting. Other researchers use parameters of some approximation of a retention curve. At least a part of the parameters in each of the models has no clear physical meaning and can only be found by fitting. The fitting models can be useful for engineering applications. However, in the best case they are limited from the viewpoint of advancement in understanding and knowledge of the links between inter- and intra-aggregate soil structure and soil water retention as a function of the structure. Therefore, nearly total domination of curve-fitting (judged by the available publications) as applied to the soil water retention seems to be unreasonable, especially in usual conditions of the spatial variability of soil hydraulic properties, when water retention modeling using only physical parameters is obviously more preferable for both data analysis and prediction. Recently the possibility of the physical prediction of another key soil characteristic, the shrinkage curve, was shown [16-21]. These works permit one to explain soil shrinkage and multiple cracking from inter- and intra-aggregate soil structure without fitting. Results of these works show that the physical (non-fitting) prediction of soil characteristics is not impossible, but is merely a difficult problem.

The objective of this work is to suggest some physical alternative to curve-fitting domination as applied to the consideration of soil water retention (drying branch) in a general case, i.e., for swell-shrink aggregated soils. This relies on the concepts and results of recent works devoted to pure clay water retention [22] (in the following referred to as the basic model) and soil shrinkage [16,18,19]. The physical model to be presented includes three parts which we consider in detail: (i) how, in a general case, a soil water retention curve can be connected with the water retention curves of a contributive clay and intra-aggregate matrix (Section **2**); (ii) the completely new version of the two-factor model of clay suction (that enables one to find the suction head of the soil including the clay as indicated in Section **2**) using the previous version [22] (the basic model) as a background for comparison (Section **3**); and (iii) the relevant aspects of the soil reference shrinkage curve model [16,18] that enable one to find the water content of the soil corresponding to its suction head (Section **4**). Then, using the model we analyze the available data in order to substantiate it (Sections **5** and **6**). Notation is summarized at the end of the paper.

*Address correspondence to this author at the Division of Environmental, Water, and Agricultural Engineering, Faculty of Civil and Environmental Engineering, Technion, Haifa 32000, Israel; E-mail: agvictor@tx.technion.ac.il

## 2. GENERAL INTERRELATIONS BETWEEN THE WATER RETENTION CURVES OF CONTRIBUTIVE CLAY, INTRA-AGGREGATE MATRIX, AND SOIL AS A WHOLE

Recently introduced new concepts relating to the intra-aggregate soil structure [16,18,19] allow one to reduce finding $h$ for an aggregated shrink-swell soil to finding $h$ for a contributive clay. The objective of this section is to show the physical links (in terms of water retention) between soil, contributive clay, and intra-aggregate matrix. These links flow out of the soil structure.

Soil volume includes aggregates that have the intra-aggregate structure (Fig.**(1)**) and inter-aggregate (structural) pores. Following [16,18] we accept that inter-aggregate pores retain their size at shrinkage, and neglect the development of possible inter-aggregate cracks (that implies the use of sufficiently small soil samples). A possible small effect of the inter-aggregate capillary cracks on soil suction can be considered separately (see e.g., [23]). The intra-aggregate structure (Fig.**(1)**) includes: (1) a deformable, but non-shrinking surface layer of aggregates (or interface layer), and (2) an intra-aggregate matrix. These two parts of aggregates have similar specific volumes and gravimetric water contents in the water-saturated state of aggregates (i.e., at maximum swelling). Both the aggregate surface layer and intra-aggregate matrix consist of a clay that surrounds silt and sand grains and, depending on soil clay content, possible so-called lacunar pores. The latter usually essentially exceed the clay matrix pores in size [24].

We start from the totally water saturated state of the soil at zero suction. In general, the maximum soil water content $W_m$ (Fig.**(2)**) consists of two contributions as

$$W_m = W_h + \Delta W_m \tag{1}$$

where $W_h$ is the maximum water content of aggregates that corresponds to the maximum swelling of the soil, aggregates, and intra-aggregate clay [16,18]; $\Delta W_m$ is the maximum water content of capillary inter-aggregate and (connected) lacunar pores (if there are any). That is, at the loss of water in the range $W_h < W < W_m$ (Fig.**(2)**), aggregates retain the water content $W_h$ and their size. Thus, the capillary suction "tail" of the soil at drying in the $W_h < W < W_m$ range (Fig.**(2)**) coincides with the capillary suction of the system of non-shrinking aggregates in the maximum swelling state. For clay soils the possible suction "tail" with $h(W_h) = h_o$ (Fig.**(2)**) (or its absence) corresponds to a horizontal section of the shrinkage curve (or its absence) at water contents higher than the maximum swelling point $W_h$ (see [16,18]). On many available experimental water retention curves, especially for soils with sufficiently high clay content, there is no suction "tail" (cf. Section **6**) because of the small value of $\Delta W = W_m - W_h \ll W_h$, (Fig.**(2)**), the small value of $h_o \ll (h - h_o)$, or inaccurate measurement. Contrary to that, for soils with sufficiently low clay content and matrix close to rigid $W_h \ll W_m$ (Fig.**(2)**), and the total water retention curve is degenerated to the "tail" only. It follows from the above that the "tail" should be considered separately in cases of special interest. Hereafter we are only interested in clay soil suction in the range $W < W_h$ (Fig.**(2)**) where the soil water content is reduced to that of aggregates, and the soil suction $h$ (we consider $h_o = 0$) is only determined by the intra-aggregate structure (Fig.**(1)**).

Below we show how the intra-aggregate structure (Fig.**(1)**) connects water retention of the soil as a whole, $h(W)$ at $W < W_h$ (Fig.**(2)**) and water retention of the clay contributing to the soil. Curve 1 in Fig.**(3)** shows a clay water retention curve, $h(\overline{w})$ where $\overline{w}$ is the water content of the clay, $\overline{w}_h$ is the maximum swelling and zero suction point of the clay. The suction of the intra-aggregate matrix including the clay (Fig.**(1)**; aggregates without surface layer) coincides with that of the clay (at a given $\overline{w}$), and water content of the intra-aggregate matrix, $w$ can be written as

$$w = c\overline{w} \tag{2}$$

where $c$ is the clay content of the soil (by weight). Curve 2 in Fig.**(3)** shows the corresponding water retention curve of an intra-aggregate matrix, $h(w)$ and corresponding axis of water content $w$ with changed scale compared with the $\overline{w}$ axis. The maximum swelling point, $w_h$ on the $w$ axis (Fig.**(3)**) corresponds to the $\overline{w}_h$ point on the $\overline{w}$ axis (about points $\overline{w}_s$ and $w_s$, see below). To find the water retention of the soil, it is sufficient to note that the suction in the soil, intra-aggregate matrix, and aggregate surface layer (Fig.**(1)**) is similar (at a given $w$), and according to [16,18] the soil water content $W$ is a sum of contributions of the aggregate surface layer, $\omega$ and intra-aggregate matrix, $w'$ as

$$W = \omega + w', \quad 0 < \omega < \omega_h, \quad 0 < w' < w'_h, \quad 0 < W < W_h \tag{3}$$

($\omega_h$, $w'_h$, and $W_h$ correspond to the maximum swelling point). With that $\omega=\omega(w')$ and $w'$ is simply connected with $w$ as

$$w'=w/K \tag{4}$$

where $K>1$ is the ratio of an aggregate solid mass to that of an intra-aggregate matrix - a new soil characteristic [16,18,19]. Equation **(4)** flows out of definitions of $w$ and $w'$. The $w$ and $w'$ values present the same water of the intra-aggregate matrix (i.e., aggregates without surface layer; Fig.**(1)**), but per unit solid mass of the matrix itself ($w$) and the soil as a whole ($w'$) (including the surface layer solid mass).

Accounting for Eq.**(3)** and **(4)** (and $\omega=\omega(w')$) we come to the conclusion that the soil suction curve, $h(W)$ (Fig.**(3)**, curves 3 and 4) is obtained from the intra-aggregate matrix curve, $h(w)$ (Fig.**(3)**, curve 2) by changing the scale along the water content axis as

$$W=\omega(w/K)+w/K, \quad 0<w<w_h, \quad\quad 0<W<W_h \;. \tag{5}$$

Together with that (see Fig.**(3)**) [16,18]

$$W_h=\omega(w_h/K)+w_h/K=w_h \tag{6}$$

because at maximum swelling ($w=w_h$) the intra-aggregate matrix and aggregate surface layer have a similar pore structure. Two possible variants of the $h(W)$ curve (Fig.**(3)**, curves 3 and 4) originate from two possible variants of the pore structure of the aggregate surface layer in the vicinity of the maximum swelling point and two corresponding dependences $\omega(w')$ ($\omega_1$ and $\omega_2$) [16,18,19]. Points $W_s=w_s/K$ and $w_s$ on the $W$ and $w$ axes, respectively, and on the $h(W)$ (Fig.**(3)**, curves 3 and 4) and $h(w)$ (Fig.**(3)**, curve 2) curves, respectively, correspond to the end point of structural shrinkage (and the initial point of basic shrinkage) when water of the aggregate surface layer exhausts, i.e., $\omega\to 0$ and $w'\to w'_s=W_s$ [16,18].

According to the above the soil water retention curve, $h(W)$ in the range $W<W_h$ (Fig.**(3)**, curves 3 and 4) is obtained from the contributive-clay water retention curve, $h(\overline{w})$ (Fig.**(3)**, curve 1) by changing the scale along the water content axis as (see Eq.**(3)** and **(6)**)

$$W=\omega(c\overline{w}/K)+c\overline{w}/K, \quad 0<\overline{w}<\overline{w}_h, \quad\quad 0<W<W_h \;. \tag{7}$$

One can also say that the $h=h(\overline{w})$ and $W=W(\overline{w})$ dependences in the range $0<\overline{w}<\overline{w}_h$ (see curve 1 in Fig.**(3)** and Eq.**(7)**) give a parametric presentation of $h(W)$ in the range $0<W<W_h$.

Thus, prediction of the soil water retention curve $h(W)$ in the range $W<W_h$ (Fig.**(3)**) is reduced to two steps. The first is the finding of the water retention curve of the contributive clay, $h=h(\overline{w})$ at $0<\overline{w}<\overline{w}_h$ (Fig.**(3)**, curve 1). The second is the finding of the relation $W=W(\overline{w})$ (see Eq.**(7)**) between the water content of contributive clay, $\overline{w}$ and that of the soil as a whole, $W$, with the following transformation of $h(\overline{w})$ to $h(W)$ at $0<W<W_h$ by changing the scale along the water content axis as $\overline{w}\to W=W(\overline{w})$. Such a two-step procedure was described above in general features (Fig.**(3)**). The water retention curve of the clay and its transformation to that of the soil by changing the scale along the water content axis as well as the necessary physical parameters are considered in detail in the following Sections. The first step (finding $h=h(\overline{w})$) is discussed in Section **3**, the second (finding $W=W(\overline{w})$) in Section **4**. The points connected to input physical parameters are considered in Section **5**.

## 3. TWO-FACTOR MODEL OF CLAY SUCTION

### 3.1. General Frames of the Model

The aim of this section is to make some preliminary notes about the structure of the model and emphasize, on the one hand its link with, and on the other hand, essential development compared to the basic model [22]. Following the basic model we present clay suction, $h$ as the product of two factors, $H$ and $Q$ (Fig.**(4)**)

$$h=HQ \;. \tag{8}$$

However, unlike the basic model, $h$ is considered in the more exact range $\zeta_a\le\zeta\le\zeta_h$ where $\zeta$ is the relative clay water content (the ratio of the current water content to the maximum one at the liquid limit); $\zeta_a$ corresponds to a boundary state when the capillary water is exhausted, but adsorbed film has the maximum thickness, $l_a$; $\zeta_h$ corresponds to the maximum swelling point. $\zeta_a$ and $\zeta_h$ are considered below (see Section **3.2**).

In addition to the statements of the basic model [22], the physical meaning of $H$ and $Q$ can be commented on as follows. The $H$ factor accumulates the effects of clay capillarity, adsorbed film, and shrinkage (see Sections **3.3-3.8**). However the capillarity is the major effect that is determined by the water configuration in the clay pore space at a given water content $\zeta_a \leq \zeta \leq \zeta_h$. Adsorbed film (of constant thickness $l_a$; see Section **3.7**) influences $H$ in the $\zeta_a \leq \zeta \leq \zeta_h$ range through its variable contribution to the total water content, $\zeta$ (that also includes capillary water). The major effect of the adsorbed water film appears in the area of very small water contents, $0 \leq \zeta \leq \zeta_a$ where the film thickness is variable. This area is not considered in this work. The shrinkage influences $H$ through decreasing pore sizes. The major effect of clay shrinkage (on $h$) is connected with rearrangement of clay particles, and this effect is accumulated in the dimensionless $Q$ factor (see Sections **3.10-3.11**).

According to the basic model, the $H$ factor for a clay only depends on one characteristic pore tube size, $R(\zeta)$ (see Sections **3.3-3.4**). However, unlike the basic model, in this work we also account for the dependence of $H$ for clay on the adsorbed water film thickness, $l_a$. Finding $H$ in this case in all the range $\zeta_a \leq \zeta \leq \zeta_h$ is one of the major tasks (see Sections **3.3-3.9**).

According to the basic model, $Q=Q(v(\zeta))$ in the range $\zeta_a \leq \zeta \leq \zeta_h$, where $v(\zeta)$ is the clay shrinkage curve in terms of the relative clay volume (the ratio of the current volume to the maximum clay volume at the liquid limit) and relative water content [25, 26]. In addition, $Q(\zeta_h)=0$ (since $h(\zeta_h)=0$, but $H(\zeta_h)\neq 0$) and $Q(\zeta)=1$ at $\zeta_a \leq \zeta \leq \zeta_z$ (Fig.(**4**); $\zeta_z$ is the clay shrinkage limit; $\zeta_z \geq \zeta_a$). Unlike the basic model where $Q$ was used in the form of the simplest approximation, below we regard the finding of $Q$ from a stricter physical consideration. Finding the $Q$ factor in the range $\zeta_a \leq \zeta \leq \zeta_h$ is one of the major tasks (see Sections **3.10-3.11**).

All developments compared to the basic model that were not indicated in this section will be noted in due course.

**3.2. Clay Water Content Range**

In the basic model $h(\zeta)$ is considered in the range $\zeta_a < \zeta < \zeta_M = 1$ that is connected with three approximations (in [22] $\zeta_a$ is designated by $\zeta_*$):

(i) the basic model neglects the difference between the maximum swelling point of the clay, $\overline{w}_h$ and the clay liquid limit, $\overline{w}_M$. This approximation, $\overline{w}_h \cong \overline{w}_M$, i.e., $\zeta_h \cong \zeta_M = 1$ differs from the real case when $\overline{w}_h < \overline{w}_M$ and $\zeta_h < 1$;
(ii) the basic model neglects the effects of the adsorbed water film; and
(iii) in force of approximation (ii) the $\zeta_a$ value is only roughly estimated to be $\sim 0.1\zeta_z$.
In this work the above approximations are removed.
(i) The interrelation between $\overline{w}_h$ and $\overline{w}_M$ for clay as

$$\overline{w}_h \cong 0.5\overline{w}_M \qquad (\zeta_h = \overline{w}_h / \overline{w}_M \cong 0.5) \qquad (9)$$

was obtained recently [16,27].
(ii) The effect of the adsorbed water film is considered in this work at such water contents where the contribution of the capillary water is more than zero, and the thickness $l_a$ of the adsorbed film is constant (see Section **3.5**). The $l_a$ thickness and summary perimeter $L(\zeta)$ of the pore tubes (per unit surface area of their cross-section) containing only the adsorbed film of the thickness $l_a$, will be considered in Sections **3.7** and **3.8** (note that we are speaking here about the pore tubes, meaning the pore channels of any cross-section shape).
(iii) The exact boundary $\zeta=\zeta_a$ (instead of the approximate $\sim 0.1\zeta_z$ value) of the water content range, where the adsorbed film thickness, $l_a$ is constant, is determined by $l_a$ and the maximum $L$ value ($L_a=L(\zeta_a)$) that corresponds to the total loss of capillary water (see Sections **3.7** and **3.8**).

Thus, below we consider the water content range $\zeta_a \leq \zeta \leq \zeta_h$ (the small area $0 \leq \zeta \leq \zeta_a$ without capillary water and with variable adsorbed-water-film thickness, $l<l_a$ is beyond the scope of this work).

**3.3. The General Expression for the $H$ Factor**

The $H$ factor (see Eq.(**8**) and Fig.(**4**)) for clay is regarded in the specified range $\zeta_a \leq \zeta \leq \zeta_h < 1$ (see Section **3.2**) where according to the basic model in a good approximation $H$ is presented as

$H = 4\Gamma\cos\alpha_c/R(\zeta), \qquad \zeta_a \leq \zeta \leq \zeta_h$. (10)

Here $\Gamma$ is the surface tension of water; $\alpha_c$ is a contact angle; and $R(\zeta)$ is a characteristic internal size of pore tubes (of any tube cross-section shape) of the clay matrix at a cross-section.

Again, according to the basic model, but in the specific range $\zeta_a \leq \zeta \leq \zeta_h$ the $R(\zeta)$ size is written as (Fig.**(5)**)

$$R(\zeta) = \begin{cases} \rho'_m(\zeta), & \zeta_n < \zeta \leq \zeta_h \\ \rho'_c(\zeta), & \zeta_a \leq \zeta \leq \zeta_n \end{cases}, \tag{11}$$

where $\rho'_m(\zeta)$ (Fig.**(5)**, curve 2) is the maximum *internal* size of pore-tube cross-sections in the $\zeta_n < \zeta \leq \zeta_h$ range ($\zeta = \zeta_n$ corresponds to the clay air-entry point); $\rho'_c(\zeta)$ (Fig.**(5)**, curve 4) is the maximum *internal* size of the water-containing pore tubes in the $\zeta_a \leq \zeta \leq \zeta_n$ range. Note that the $H$ presentation in Eq.**(10)** reflects the physical peculiarity of a clay matrix structure. There is only one independent characteristic size, $R(\zeta)$ (unlike in the general case of soil). Indeed, at least in the area of normal (or basic) shrinkage, $\zeta_n < \zeta \leq \zeta_h < 1$ there is only one characteristic size - the maximum internal size of pore-tube cross-sections $\rho'_m(\zeta)$ (Fig.**(5)**, curve 2), that coincides with the maximum *internal* size $\rho'_f(\zeta)$ of the water-filled pore tubes in this area.

Note, that in this work, unlike in the basic model, the size $R(\zeta)$ entering Eqs.**(10)** and **(11)**, in addition to finding in the specified range, is determined as affected by the adsorbed water film (see Sections **3.5-3.9**).

### 3.4. Expression for the Characteristic Pore Tube Size (*R*) of Saturated Clay Matrix

According to the basic model in the area $\zeta_n \leq \zeta \leq \zeta_h$ where the clay is in the saturated state (i.e., without air in pores) $R(\zeta) \cong \rho'_m(\zeta)$ (Eq.**(11)**; Fig.**(5)**, curve 2) is expressed through $v(\zeta)$, $v_z$ ($v_z \equiv v(\zeta_z)$ is the $v$ value at the shrinkage limit of the clay, $\zeta = \zeta_z$), $v_s$ (the relative volume of clay solids, i.e., the ratio of the solid volume to clay volume at the liquid limit), $r_{mM}$ (the maximum external size of clay pores at $\zeta = 1$); and characteristic constants of the clay microstructure, $\alpha \cong 1.41$ and $A \cong 13.57$ [25] as

$$R(\zeta) = \rho'_m(\zeta) = \rho_m(\zeta) - \delta(\zeta) = r_{mM} v(\zeta)^{1/3} [0.75 - (\alpha/A)(v_s/v(\zeta))], \qquad \zeta_n \leq \zeta \leq \zeta_h \tag{12}$$

where $\rho_m(\zeta)$ is the maximum *external* size of clay pore-tube cross-sections (i.e., the size including the half thickness of clay particles that outline the pores), and $\delta(\zeta)$ is the mean thickness of clay particle cross-sections. Note, that $R(\zeta)$ at $\zeta_n \leq \zeta \leq \zeta_h$ (Fig.**(5)**, curve 2) from Eq.**(12)** does not depend explicitly on the water content $\zeta$, but only through the shrinkage curve $v(\zeta)$. That is, $R(\zeta) = \rho'_m(\zeta) = \rho'_m(v(\zeta))$ at $\zeta_n \leq \zeta \leq \zeta_h$.

### 3.5. Equation for the Characteristic Pore Tube Size (*R*) of an Unsaturated Clay Matrix

Unlike $R$ in the $\zeta_n \leq \zeta \leq \zeta_h$ range (see Section **3.4**) the characteristic size $R(\zeta) = \rho'_c(\zeta)$ at $\zeta_a \leq \zeta \leq \zeta_n$ (see Eq.**(11)**; Fig.**(5)**, curve 4) cannot be written immediately, since in this range the pore water configuration is more complex. In addition to the water-filled pore tubes (with maximum internal size $\rho'_f(\zeta)$; Fig.**(5)**, curve 3) there are also the water-containing pore tubes (with maximum internal size $\rho'_c(\zeta)$; Fig.**(5)**, curve 4) with air, capillary water, and adsorbed water, as well as the pore tubes with only adsorbed water film of maximum thickness $l_a$ (with maximum internal size $\rho'_m(\zeta)$; Fig.**(5)**, curve 1). For such a water configuration in the clay at a given water content $\zeta_a \leq \zeta \leq \zeta_n$ one can write the water balance equation that accounts for the corresponding contributions to the total water content. The *unknown* $\rho'_c(\zeta)$ characteristic size in the area, $\zeta_a \leq \zeta \leq \zeta_n$ (Eq.**(11)**; Fig.**(5)**, curve 4) enters this equation and can be found as its solution (see Section **3.9**). The objective of this section is only to present and comment on this equation. The water balance equation (at a clay cross-section) can be written as

$$F(\zeta) = \varphi(\rho'_f(\zeta)) + \int_{\rho'_f(\zeta)}^{\rho'_c(\zeta)} g(\rho') \frac{d\varphi}{d\rho'} d\rho' + l_a L(\rho'_c(\zeta), \zeta), \qquad \zeta_a \leq \zeta \leq \zeta_n \tag{13}$$

where $F$ is the pore volume fraction occupied by water (or saturation degree), at a relative water content $\zeta$; $\varphi(\rho')$ is the pore tube-size distribution, in particular, $\varphi(\rho'_f)$ is the $\varphi$ value at $\rho' = \rho'_f$; $g(\rho')$ is the degree of filling with water of the pore tubes of internal $\rho'$ size ($0 < g < 1$) if $\rho'_f < \rho' < \rho'_c$; and $L(\rho'_c(\zeta), \zeta)$ is the summary perimeter of the pore tubes (per unit surface area of their cross-section) containing only the adsorbed film of thickness $l_a$ at a given water content $\zeta$ and corresponding *unknown* $\rho'_c(\zeta)$ value. The first and second terms in the right part of Eq.**(13)** give the contributions of the water-filled and water-containing pore tubes, respectively, to the total water content. The third term gives the contribution of the adsorbed film of pore tubes without capillary water. For the clay with a shrinkage curve $v(\zeta)$ the saturation degree $F(\zeta)$ in Eq.**(13)** is found to be [25]

$$F(\zeta) = [(1 - v_s)/(v(\zeta) - v_s)]\zeta, \qquad 0 < \zeta < 1 \ . \tag{14}$$

It is obvious that in the boundary states of the $\zeta_a \leq \zeta \leq \zeta_n$ range, the right part of the Eq.(13) is only reduced to the contribution of adsorbed water [$l_a L(\rho'_c(\zeta_a),\zeta_a)$] at $\zeta=\zeta_a$ or that of the water filled pores [$\varphi(\rho'_f(\zeta_n))$] at $\zeta=\zeta_n$. However, before solving Eq.(13) we should consider the first (Section **3.6**), second (Section **3.9**), and third (Sections **3.7** and **3.8**) terms in the right part of Eq.(13) in more detail. Finally, note that a similar equation has been used in the basic model, but without the adsorbed-water term. In addition, the two first terms in the right part of Eq.(13) will be essentially modified compared with the basic model (see Sections **3.6** and **3.9**).

**3.6. The Pore Tube-Size Distribution of a Clay Matrix**

The presentation form of a pore-size distribution plays an important role. Chertkov [25,26] used the presentation that is convenient for considering clay shrinkage. The convenience consists in the use of an *external* pore ($r$) and pore tube ($\rho$) sizes (i.e., the sizes that include a half-thickness of clay particles limiting the pores). In this case the volume of any pore, that is proportional to $r^3$, is proportional to the clay volume $V$ at shrinkage. However, such a presentation does not include, in an explicit form, the clay porosity that is connected with *internal* pore sizes ($r'$ and $\rho'$) which determine the clay water retention. For this reason the use of the pore-size distribution presentation from [25, 26] in the basic model [22] required special normalization of the distribution. The generalization, giving the more convenient presentation of the pore-size distribution, using internal pore sizes, and explicitly including porosity as a distribution parameter, was recently suggested [28]. In addition (unlike the presentation from [25, 26, 22]), this presentation is generalized in a natural way to a two- or multi-mode porosity case that can be topical for clay and soil.

We use the simplest pore-tube size distribution, $\varphi(\rho') \equiv \varphi(x(\rho'))$ for the two-dimensional situation (with one-mode porosity) from the intersecting-surfaces approach [28] as

$$\varphi(x(\rho'))=[1-(1-P)^{I(x(\rho'))/I(1)}]/P \tag{15}$$

where

$$x(\rho')=(\rho'-\rho'_{min})/(\rho'_m-\rho'_{min}), \quad 0<x\leq 1; \tag{16}$$

$$I(x)=\ln(6)(3x)^3\exp(-3x), \quad (I(1)=2.4086) \quad 0<x\leq 1 \ . \tag{17}$$

In Eq.(15) $P$ is porosity (note that both the definitions of porosity - volumetric and areal - must give the same values; see, e.g., [29]). In Eq.(16) $\rho'_{min}$ and $\rho'_m$ are the minimum and maximum pore-tube cross-section sizes, respectively.

The presentation of Eqs.(15)-(17) is relevant for both rigid and non-rigid matrices (with one-mode porosity). However, in the case of a clay matrix (a non-rigid matrix) Eqs.(15) and (16) should also be specified because the clay matrix parameters, $\rho'_{min}$, $\rho'_m$ (Eq.(16)) and $P$ (Eq.(15)) become functions of the relative water content, $\zeta$. We are interested in $\varphi(\rho')$ as well as $\rho'_{min}$, $\rho'_m$, and $P$ at $\zeta_a \leq \zeta \leq \zeta_n$ (see Eq.(13)). The expression for $\rho'_m(\zeta)$ at $\zeta_a \leq \zeta \leq \zeta_n$ (Fig.(**5**), curve 1) coincides with the expression of the same clay matrix parameter at $\zeta_n \leq \zeta \leq \zeta_h$ in Eq.(12) (see Fig.(**5**), curve 2) [22]. The clay porosity, $P(\zeta)$ was defined in [25] as

$$P(\zeta)=1-v_s/v(\zeta) \ . \qquad \zeta_a \leq \zeta \leq \zeta_n \ . \tag{18}$$

Defining the expression for $\rho'_{min}(\zeta)$ of clay as (see Fig.(**5**), curve 5)

$$\rho'_{min}(\zeta)=\rho_o(\zeta)-\delta(\zeta)=r_o(\zeta)-\Delta(\zeta)=r_{mM}(v_s/A)v(\zeta)^{1/3}[\gamma-1/v(\zeta)], \qquad 0<\zeta<1 \ . \tag{19}$$

we take that the minimum internal pore tube size ($\rho_o-\delta$) coincides with the minimum internal size of pores ($r_o-\Delta$) ($\Delta$ is the mean thickness of clay particles [25], $r_o$ [25] and $\rho_o$ [22] are the minimum external sizes of pores and pore tubes, respectively, and $\gamma \cong 9$ is the characteristic clay pore constant [25]). This definition of $\rho'_{min}(\zeta)$ (Eq.(19)) differs from that in the basic model.

Note that the $\rho'_{min}(\zeta)$ (Eq.(19)), $\rho'_m(\zeta)$ (Eq.(12)), and $P(\zeta)$ (Eq.(18)) dependences on $\zeta$ are determined by the clay shrinkage curve $v(\zeta)$ (for $v(\zeta)$ see [25,26]). In addition, it follows from the definitions and physical meaning of $\rho'_c(\zeta)$ (see Section **3.3**; Eq.(11); Fig.(**5**), curve 4), $\rho'_f(\zeta)$ (see Section **3.3**; Fig.(**5**), curve 3), and $\rho'_{min}(\zeta)$ (Eq.(19)) that

$$R(\zeta_a) \equiv \rho'_c(\zeta_a)=\rho'_f(\zeta_a)=\rho'_{min}(\zeta_a) \tag{20}$$

(see Fig.**(5)**, curves 3, 4, and 5) which will be used in Section **3.9**. Finally, note that the pore-tube size distribution from Eq.**(15)**-**(17)** does not take into account the intra-particle (or inter-layer) pores and for this reason relates to clays with negligible inter-layer porosity.

### 3.7. The Summary Perimeter of Pore Tubes ($L$) Containing only the Adsorbed Water Film

The $L$ value can be written as follows

$$L(\rho'_c(\zeta),\zeta) = 4 \int_{\rho'_c(\zeta)}^{\rho'_m(\zeta)} \frac{1}{\rho'} \frac{d\varphi}{d\rho'} d\rho', \qquad \zeta_a \leq \zeta \leq \zeta_n \qquad (21)$$

where $\varphi = \varphi(x(\rho'))$ from Eq.**(15)**; $x(\rho')$ from Eq.**(16)**; $\rho'_m(\zeta)$ from Eq.**(12)**, and $\rho'_c(\zeta)$ is so far an unknown dependence (see Section **3.5**; Fig.**(5)**, curve 4). Indeed, $(d\varphi/d\rho') d\rho'$ is the fraction of the pore-tube cross-section surface area for tubes with internal size $\rho'$ in the small range $d\rho'$. The value, $(1/\rho'^2)(d\varphi/d\rho') d\rho'$ gives the number of corresponding pore tubes, and $(4/\rho')(d\varphi/d\rho') d\rho'$ is their perimeter if their cross-section is considered to be a square or a circle. Integration in Eq.**(21)** between $\rho'_c(\zeta)$ and $\rho'_m(\zeta)$ at a given $\zeta$ (see Fig.**(5)**, between curves 4 and 1) gives the summary perimeter of pore tubes, $L(\rho'_c(\zeta),\zeta)$ containing only the adsorbed film of thickness $l_a$. Equation **(21)** can be rewritten in the form that is more suitable for numerical calculations as

$$L(\rho'_c,\zeta) = \frac{4}{\rho'_m(\zeta)} - \frac{4\varphi(x(\rho'_c))}{\rho'_{min}(\zeta) + x(\rho'_c)Z(\zeta)} + \frac{4}{Z(\zeta)} \int_{x(\rho'_c)}^{1} \frac{\varphi(x')dx'}{[M(\zeta)+x']^2}, \qquad \zeta_a \leq \zeta \leq \zeta_n \qquad (22)$$

where $Z(\zeta) \equiv \rho'_m(\zeta) - \rho'_{min}(\zeta)$ and $M(\zeta) \equiv \rho'_{min}(\zeta)/[\rho'_m(\zeta) - \rho'_{min}(\zeta)]$. This expression of $L$ is used in Section **3.9** when solving Eq.**(13)**. In addition, we need the boundary value $L_a \equiv L(\rho'_c(\zeta_a),\zeta_a) = L(\rho'_{min}(\zeta_a),\zeta_a)$ (see Eq.**(20)** and Fig.**(5)**, curves 4 and 5). $L_a$ can be found without solving Eq.**(13)** and is necessary for estimating $l_a$ (in Section **3.8**). At $\zeta = \zeta_a$ Eq.**(22)** is reduced to

$$L_a = \frac{4}{\rho'_m(\zeta_a)} + \frac{4}{Z(\zeta_a)} \int_0^1 \frac{\varphi(x')dx'}{[M(\zeta_a)+x']^2} \ . \qquad (23)$$

As noted $\rho'_m(\zeta)$ (Eq.**(12)**), $\rho'_{min}(\zeta)$ (Eq.**(19)**), and $P(\zeta)$ (Eq.**(18)**) depend implicitly on $\zeta$, only through the shrinkage curve, $v(\zeta)$. Since at $\zeta_a \leq \zeta_z$ $v_a \equiv v(\zeta_a) = v_z \equiv v(\zeta_z)$, in fact $\rho'_m(\zeta_a) = \rho'_m(\zeta_z)$, $\rho'_{min}(\zeta_a) = \rho'_{min}(\zeta_z)$ (see Fig.**(5)**), $P_a \equiv P(\zeta_a) = P_z = 1 - v_s/v_z$, and correspondingly, $L_a = L_z$ (see Eq.**(23)**). Finally, note that the dimension of all $L$ values is the reciprocal of length. We should now consider the maximum thickness $l_a$ of an adsorbed water film in clay.

### 3.8. The Maximum Thickness ($l_a$) of an Adsorbed Film and Corresponding Boundary Water Content ($\zeta_a$)

We rely on two *physical* conditions.
(i) At usual temperatures (~20°C) separate clay particles are covered by adsorbed film already before the formation of the clay particle network. For this reason, pores (and pore tubes) of a size that is less than $2l_a$ cannot appear at clay matrix formation (we mean the absence of external loads leading to clay compaction and consolidation). That is, even at $\zeta = \zeta_a$ the minimum size of pore tubes, $\rho'_{min}(\zeta_a)$ should be no less than $2l_a$ as

$$\rho'_{min}(\zeta_a) \geq 2l_a \ . \qquad (24)$$

Note that from Eq.**(19)** for a given clay (i.e., at given $r_{mM}$, $v_s$, and $v_z$, see [25]) one has $\rho'_{min}(\zeta_a) = \rho'_{min}(\zeta_z) = r_{mM}(v_s/A)v_z^{1/3}(\gamma - 1/v_z)$.
(ii) The clay water content, that corresponds to the presence of an adsorbed film only, should not exceed the water content that corresponds to the shrinkage limit. In terms of the saturation degree, $F$ (see Eq.**(14)**) that is written as follows

$$F_a \equiv F(\zeta_a) \leq F_z \equiv F(\zeta_z) \ . \qquad (25)$$

Since (by definition of $F(\zeta_a)$, $l_a$, and $L_a \equiv L_z$) $F_a = l_a L_z$ the condition of Eq.**(25)** can be rewritten as

$$l_a \leq F_z/L_z \tag{26}$$

where $F_z/L_z$ (similar to $\rho'_{min}(\zeta_a)$ in Eq.(24)) depends on the specific parameters ($r_{mM}$, $v_s$, and $v_z$) of a given clay (see Eq.(14) and Section 3.7).

In addition, it follows from the meaning of the condition leading to Eq.(24) that at clay particle network (or clay matrix) formation the size of the minimum pores (at $\zeta=\zeta_a$) strives to be maximally close to $2l_a$ and, on the contrary, $2l_a$ should be maximally close to $\rho'_{min}(\zeta_a)$. This additional condition together with Eqs.(24) and (25) implies that $2l_a=\rho'_{min}(\zeta_a)$ if only $\rho'_{min}(\zeta_a)<2F_z/L_z$. Otherwise, the same additional condition (of the maximum proximity of $2l_a$ to $\rho'_{min}(\zeta_a)$) means that $2l_a=2F_z/L_z$. Finally, one can write $2l_a$ as

$$2l_a = \min(\rho'_{min}(\zeta_a), 2F_z/L_z) \tag{27}$$

(it is worth reiterating that $\rho'_{min}(\zeta_a)$ and $2F_z/L_z$ are the known functions of $r_{mM}$, $v_s$, and $v_z$). Thus, for a given clay, depending on its specifications (usually connected with clay mineralogy and cation set) that are reflected by the $r_{mM}$, $v_s$, and $v_z$ clay matrix characteristics [25], one of two the following possibilities can be realized.
(i) If

$$\rho'_{min}(\zeta_a) < 2F_z/L_z \tag{28}$$

then (see Eq.(27))

$$2l_a = \rho'_{min}(\zeta_a) \tag{29}$$

and, correspondingly,

$$F_a = l_a L_z < F_z \tag{30}$$

and (see Eq.(14))

$$\zeta_a = [(v_z-v_s)/(1-v_s)] F_a < \zeta_z = [(v_z-v_s)/(1-v_s)] F_z \tag{31}$$

(Fig.(5) presents such a situation).
(ii) If

$$\rho'_{min}(\zeta_a) \geq 2F_z/L_z \tag{28'}$$

then (see Eq.(27))

$$l_a = F_z/L_z . \tag{29'}$$

Correspondingly,

$$F_a = l_a L_z = F_z \tag{30'}$$

and (cf.Eq.(31))

$$\zeta_a = \zeta_z \tag{31'}$$

(i.e., the points $\zeta_a$ and $\zeta_z$ in Fig.(5) merge).
Note, that irrespective of which of the two above possibilities takes place, i.e., at $l_a$ from both Eq.(29) and (29') (see Fig.(5)), the Eq.(20) is fulfilled. See estimates of $l_a$, $L_a$, $F_a$, and $\zeta_a$ for real soils in Section 6.

### 3.9. Solution of the Water Balance Equation for an Unsaturated Clay Matrix

When solving Eq.(13) with respect to $\rho'_c(\zeta)$ we use the following boundary conditions.
(i) At the boundary water content $\zeta=\zeta_a$ the maximum internal size of the water filled ($\rho'_f(\zeta)$; Fig.(5), curve 3) and water containing ($\rho'_c(\zeta)$; Fig.(5), curve 4) pore-tube cross-sections should coincide (this condition was not used in the basic model)

$\rho'_f(\zeta_a) = \rho'_c(\zeta_a)$ . (32)

(ii) At the boundary water content $\zeta=\zeta_n$ $\rho'_f(\zeta)$ (Fig.**(5)**, curve 3) and $\rho'_c(\zeta)$ (Fig.**(5)**, curve 4) also coincide

$\rho'_f(\zeta_n) = \rho'_c(\zeta_n)$ . (33)

(iii) $R(\zeta)$ (Eq.**(11)**) should be smooth at $\zeta=\zeta_n$ (Fig.**(5)**). That is,

$\rho'_c(\zeta_n) = \rho'_m(\zeta_n)$   and   $d\rho'_c(\zeta)/d\zeta|_{\zeta=\zeta_n} = d\rho'_m(\zeta)/d\zeta|_{\zeta=\zeta_n}$ . (34)

In addition, the $\rho'_f(\zeta)$ and $\rho'_c(\zeta)$ functions (Fig.**(5)**, curves 3 and 4, respectively) should meet the obvious physical condition (which was not used in the basic model) that the water-containing (i.e., non-totally filled) pore tubes give a small contribution to the water balance equation (Eq.**(13)**). That is, the second term in the right side of Eq.**(13)** is small as

$$\varphi(\rho'_f(\zeta)) + l_a L(\rho'_c(\zeta),\zeta) \gg \int_{\rho'_f(\zeta)}^{\rho'_c(\zeta)} g(\rho') \frac{d\varphi}{d\rho'} d\rho', \qquad \zeta_a \le \zeta \le \zeta_n \qquad (35)$$

It follows that independently of an exact form of $g(\rho')$ dependence, $\rho'_c(\zeta)$ differs from $\rho'_f(\zeta)$ by the small addition, $\delta\rho'_f$ as (Fig.**(5)**)

$\rho'_c = \rho'_f + \delta\rho'_f \qquad \zeta_a \le \zeta \le \zeta_n$ (36)

where $\delta\rho'_f(\zeta) \ll \rho'_f(\zeta)$ and according to Eq.**(32)** and **(33)**

$\delta\rho'_f(\zeta_a) = \delta\rho'_f(\zeta_n) = 0$ . (37)

Then (without additional assumptions with respect to the $g(\rho')$ function and pore shape as in the basic model) we can find $\rho'_c(\zeta)$ in the range $\zeta_a < \zeta < \zeta_n$ in the first approximation as $\rho'_c(\zeta) = \rho'_f(\zeta)$ where $\rho'_f(\zeta)$ (Fig.**(5)**, curve 3) is the solution of the equation

$F(\zeta) = \varphi(\rho'_f) + l_a L(\rho'_f,\zeta)$, $\qquad \zeta_a \le \zeta \le \zeta_n$ . (38)

Note that the left side of Eq.**(38)** depends only on $\zeta$ (Eq.**(14)**), but the right side (see Eqs.**(15)**-**(17)**) depends on both $\zeta$ (through $\rho'_{min}(\zeta)$, $\rho'_m(\zeta)$, and $P(\zeta)$ of Eqs.**(19)**, **(12)**, and **(18)**, respectively) and $\rho'_f$. Thus, at a given $\zeta_a \le \zeta \le \zeta_n$ one can (numerically) find the corresponding $\rho'_f$ value, i.e., $\rho'_c(\zeta) \equiv \rho'_f(\zeta)$ dependence in the first approximation.

The simplest way to find $\rho'_c(\zeta)$ in the second approximation is as follows. One can write $\rho'_c(\zeta)$ at $\zeta_a \le \zeta \le \zeta_n$ (Fig.**(5)**, curve 4) as

$$\rho'_c(\zeta) = \begin{cases} \rho'_f(\zeta), & \zeta_a \le \zeta \le \zeta' \\ \rho'_f(\zeta) + \delta\rho'_f(\zeta), & \zeta' < \zeta \le \zeta_n \end{cases}. \qquad (39)$$

It follows from the general qualitative picture (Fig.**(5)**) that $\max(\delta\rho'_f) = \max(\rho'_c - \rho'_f)$ is reached close to $\zeta=\zeta_n$. That is, the point of "sewing" $\zeta=\zeta'$ (Fig.**(5)**; Eq.**(39)**) is also close to $\zeta=\zeta_n$ (i.e., $\zeta_n - \zeta' \ll \zeta' - \zeta_a$; that is confirmed by direct calculations, see Section **6**). For this reason we approximate $\rho'_c = \rho'_f + \delta\rho'_f$ at $\zeta' < \zeta \le \zeta_n$ to be

$\rho'_c(\zeta)/\rho'_m(\zeta) = 1 - G(\zeta - \zeta_n)^2$, $\qquad \zeta' < \zeta \le \zeta_n$ . (40)

Then conditions at $\zeta=\zeta_n$ (Fig.**(5)**) given by Eq.**(34)** are obviously fulfilled. The $G$ coefficient and "sewing" point $\zeta=\zeta'$ (Fig.**(5)**) are found from conditions of the smooth connection between $\rho'_c(\zeta) = \rho'_f(\zeta)$ from Eq.**(39)** and $\rho'_c(\zeta)$ from Eq.**(40)** at $\zeta=\zeta'$. The $\rho'_c(\zeta)$ found in good approximation meets Eq.**(13)** and conditions from Eq.**(32)**-**(36)** at $\zeta_a \le \zeta \le \zeta_n$ (Fig.**(5)**).

It is worth emphasizing an important property of $R(\zeta) = \rho'_c(\zeta) \equiv \rho'_f(\zeta) + \delta\rho'_f(\zeta)$ at $\zeta_a \le \zeta \le \zeta_n$. In this range (unlike the $\zeta_n \le \zeta \le \zeta_h$ range, see the end of Section **3.4**) $\rho'_f(\zeta) = \rho'_f(\zeta,\nu(\zeta))$. Hence, the similar presentation also relates to $\rho'_c(\zeta)$ and $R(\zeta)$. That is, in this range $R(\zeta) = R(\zeta,\nu(\zeta))$. Indeed, the right side of Eq.**(38)** (besides $\rho'_f$) only implicitly depends on $\zeta$ through $v(\zeta)$ (because $\rho'_{min}$, $\rho'_m$, and $P$ depend on $\zeta$ through $v(\zeta)$), but the left side of Eq.(**38**) depends on $\zeta$ both explicitly and through $v(\zeta)$ (see Eq.(**14**)). Finally, calculation of $R(\zeta)$ (Eq.(**11**); Fig.(**5**), curves 2 and 4) together with Eq.(**10**) gives $H(\zeta)$. Hence, the $H(\zeta)$ dependence (similar to $R(\zeta)$) has the following $\zeta$ structure

$$H(\zeta) = \begin{cases} H(\zeta, v(\zeta)), & \zeta_a \leq \zeta \leq \zeta_n \\ H(v(\zeta)), & \zeta_n \leq \zeta \leq \zeta_h \end{cases}. \tag{41}$$

### 3.10. Estimating the *Q* Factor Dependence on the Relative Water Content of a Clay

The simple approximate presentation of $Q$ in the basic model should be replaced with a more exact consideration. At $\zeta_a \leq \zeta \leq \zeta_z$ $Q(v(\zeta))=1$ (see Section **3.1** and Fig.(**4**)). We are interested in the $Q(v(\zeta))$ behavior at $\zeta_z \leq \zeta \leq \zeta_h$ (Fig.(**4**)). Since $\zeta_h - \zeta_z < 1$ it is reasonable to present $Q(\zeta)$ in this area as an expansion in powers of $(\zeta - \zeta_z)$ or $(\zeta_h - \zeta)$.

Besides the physically distinguished points, $\zeta_z$ (the shrinkage limit) and $\zeta_h$ (the maximum swelling point) in the range $\zeta_z \leq \zeta \leq \zeta_h$ there is yet one other physically distinguished point, $\zeta_n$ (Fig.(**4**); the air-entry point) in which the character of the clay shrinkage curve changes (the curve becomes nonlinear). Therefore, it is natural to divide the $\zeta_z \leq \zeta \leq \zeta_h$ range into two smaller ones, $\zeta_z \leq \zeta \leq \zeta_n$ and $\zeta_n \leq \zeta \leq \zeta_h$ (Fig.(**4**)) and present $Q(\zeta)$ in the former as an expansion in powers of $(\zeta - \zeta_z)$ and in the latter as an expansion in powers of $(\zeta_h - \zeta)$.

One should remember that $Q$ depends on $\zeta$ through the shrinkage curve $v(\zeta)$, $Q(\zeta)=Q(v(\zeta))$, and because we use for $v(\zeta)$ at $\zeta_z \leq \zeta \leq \zeta_n$ the approximation connected with the expansion in powers of $(\zeta - \zeta_z)$ up to the second power $\sim (\zeta - \zeta_z)^2$ [25,26,22], the expansions for the $Q(\zeta)$ should also be limited by the squared approximation as

$$Q(\zeta)=q_1+q_2(\zeta-\zeta_z)+q_3(\zeta-\zeta_z)^2, \qquad \zeta_z \leq \zeta \leq \zeta_n \tag{42a}$$

$$Q(\zeta)=q_4+q_5(\zeta_h-\zeta)+q_6(\zeta_h-\zeta)^2, \qquad \zeta_n \leq \zeta \leq \zeta_h. \tag{42b}$$

Accounting for the obvious boundary conditions $Q(\zeta_z)=1$, $Q'(\zeta_z)=0$, and $Q(\zeta_h)=0$ (see Fig.(**4**)) Eq.(**42**) is reduced to

$$Q(\zeta)=1-Q_1(\zeta-\zeta_z)^2, \qquad \zeta_z \leq \zeta \leq \zeta_n \tag{43a}$$

$$Q(\zeta)=Q_2(\zeta_h-\zeta)+Q_3(\zeta_h-\zeta)^2, \qquad \zeta_n \leq \zeta \leq \zeta_h \tag{43b}$$

($q_3 \equiv -Q_1$, $q_5 \equiv Q_2$, and $q_6 \equiv Q_3$ were re-denoted for convenience).

Since $Q(\zeta)$ is at its maximum at $\zeta=\zeta_z$ (Fig.(**4**)), $Q_1$ in Eq.(**43a**) should be positive, $Q_1>0$. In addition the $Q_2$ and $Q_3$ coefficients in Eq.(**43b**) should be non-negative, $Q_2 \geq 0$ and $Q_3 \geq 0$. This stems from the considerations about the existence and position of the inflexion point of $Q(\zeta)$. Indeed, the shrinkage curve, $v(\zeta)$ inflexion point (in any approximation) is at $\zeta=\zeta_n$ [25,26]. Since $Q(\zeta)=Q(v(\zeta))$, the $Q$ factor inflexion point should also be at $\zeta=\zeta_n$. Accounting for the negative curvature ($-Q_1$) of $Q(\zeta)$ (Eq.(**43a**)) at $\zeta_z \leq \zeta \leq \zeta_n$ (Fig.(**4**)) it follows that $Q(\zeta)$ from Eq.(**43b**) should have the non-negative curvature, $Q_3 \geq 0$ (Fig.(**4**)). In addition, since $Q(\zeta)$ should monotonously decrease with $\zeta$ increase, $Q_2 \geq 0$. These conditions, $Q_1>0$, $Q_2 \geq 0$ and $Q_3 \geq 0$ will be used below.

Thus, we have three coefficients, $Q_1$, $Q_2$ and $Q_3$ and two "sewing" conditions at $\zeta=\zeta_n$

$$Q(\zeta_{n-})=Q(\zeta_{n+}) \tag{44a}$$

$$Q'(\zeta_{n-})=Q'(\zeta_{n+}) \tag{44b}$$

($\zeta_{n-}$ and $\zeta_{n+}$ correspond to approaching to $\zeta=\zeta_n$ from the left and right, respectively). The physical meaning of these conditions is the smoothness of $Q(\zeta)$ at $\zeta=\zeta_n$ (Fig.(**4**)). It is worth emphasizing that a condition similar to Eq.(**44b**), but for the second derivatives of $Q(\zeta)$ at $\zeta=\zeta_n$, does not take place in the squared approximation used. Indeed, the second derivative of $v(\zeta)$ is subject to a break at $\zeta=\zeta_n$ [25,26]. Hence, $Q''(\zeta_{n-}) \neq Q''(\zeta_{n+})$. Using Eqs.(**44**) the $Q_2$ and $Q_3$ coefficients can be expressed through $Q_1$. However, for the following instead of $Q_1$ we introduce the parameter $Q_n \equiv Q(\zeta_n)$ (see Fig.(**4**)) with more immediate physical meaning. According to Eq.(**43a**) at $\zeta=\zeta_n$

$$Q_1=(1-Q_n)/(\zeta_n-\zeta_z)^2. \tag{45}$$

Equations **(43)** and **(44)**, after the replacement of $Q_1$ with $Q_n$ from Eq.**(45)**, give

$$Q_2=2Q_n/(\zeta_h-\zeta_n)-2(1-Q_n)/(\zeta_n-\zeta_z) ,\tag{46}$$

$$Q_3=[2(1-Q_n)(\zeta_h-\zeta_n)/(\zeta_n-\zeta_z)-Q_n]/(\zeta_h-\zeta_n)^2 .\tag{47}$$

Here $Q_n$ (see Fig.**(4)**) is the (so far unknown) physical parameter that determines the coefficients $Q_1$ (Eq.**(45)**), $Q_2$ (Eq.**(46)**), and $Q_3$ (Eq.**(47)**) in Eq.**(43)** and thereby the $Q(\zeta)$ factor for a given clay.

Similar to $\zeta_z$ and $\zeta_n$ [25,26] ($\zeta_h=0.5$ [16]), $Q_n$ can be expressed through the clay characteristics, $v_s$ and $v_z$ [25,26]. The dependence $Q_n(v_s,v_z)$ is discussed in the following Section. However, before the consideration of $Q_n(v_s,v_z)$ the obvious range, $0<Q_n<1$ (see Fig.**(4)**) (that also flows out of Eq.**(45)** and $Q_1>0$) can be essentially diminished. Indeed, the above conditions, $Q_2 \geq 0$ and $Q_3 \geq 0$, using Eqs.**(46)** and **(47)**, reduce the range where the $Q_n(v_s,v_z)$ value can be to

$$0<Q_{nmin} \leq Q_n \leq Q_{nmax}<1 ,\tag{48}$$

where

$$Q_{nmin}=(\zeta_h-\zeta_n)/(\zeta_h-\zeta_z),\tag{49a}$$

$$Q_{nmax}=2(\zeta_h-\zeta_n)/[2(\zeta_h-\zeta_n)+(\zeta_n-\zeta_z)] .\tag{49b}$$

Accounting for $\zeta_z<\zeta_n<\zeta_h$ (see Fig.**(4)**) one can see that the obvious condition $Q_{nmin}<Q_{nmax}$ is always fulfilled. The effect of a possible $Q_n$ from Eqs.**(48)** and **(49)** on $Q(\zeta)$ dependence is illustrated in Fig.**(4)**. Checking the experimental feasibility of the clay $Q(\zeta,Q_n)$ factor presentation by Eqs.**(43)**, **(45)**-**(47)** using the $Q_n$ value from Eqs.**(48)** and **(49)** is considered in Sections **5** and **6**.

### 3.11. Estimating the $Q_n$ Parameter for a Clay through the Clay Porosity at the Air-Entry Point ($P_n$)

We want to express $Q_n$ through a value that can be easily computed for a given clay, i.e., through clay parameters, $v_s$ and $v_z$. Then, knowing this value we will be able to calculate $Q_n$ for the clay. By definition $Q_n=Q(\zeta_n)$ where $\zeta_n=(v_n-v_s)/(1-v_s)$, $v_n=v_s/(1-P_n)$ [25,26], and $P_n$ is the clay porosity at the air-entry point. Thus, $Q_n=Q_n(P_n)$, and $Q_n(P_n)$ should be a universal function that is applicable to any clay.

The $Q_n(P_n)$ dependence can be characterized as follows. For clays (contributing the real soils) $P_n$ and $Q_n$ vary in the ranges

$$0.3 \cong P_{n\,low}<P_n<P_{n\,up} \cong 0.8 ,\tag{50a}$$

$$0.5 \cong Q_{n\,low}<Q_n<Q_{n\,up} \cong 0.9 ,\tag{50b}$$

where $P_{n\,low}$ and $P_{n\,up}$ are the lower and upper boundaries of $P_n$, and $Q_{n\,low}$ and $Q_{n\,up}$ are the similar boundaries of $Q_n$ variations (see Fig.**(6)**). With that $Q_n$ increases with $P_n$ decrease and vice versa (Fig.**(6)**). Indeed, the $P_n$ decrease means the transition to a more rigid clay with larger $Q_n$. Hence

$$Q_n \rightarrow Q_{n\,up} \quad \text{at} \quad P_n \rightarrow P_{n\,low} .\tag{51}$$

In addition (see Fig.**(6)**)

$$Q_{n\,up}-Q_n(P_n)<<P_n-P_{n\,low} \qquad \text{at} \quad P_n \rightarrow P_{n\,low} .\tag{52}$$

That is, $Q_n$ grows very slowly when $P_n \rightarrow P_{n\,low}$. For this reason we take (see Fig.**(6)**)

$$Q_n'(P_n) \rightarrow 0 \text{ and } Q_n''(P_n) \rightarrow 0 \qquad \text{at} \quad P_n \rightarrow P_{n\,low} .\tag{53}$$

For convenience we introduce the relative values, $p$ and $q$ as

$$p=(P_n-P_{n\,low})/(P_{n\,up}-P_{n\,low}) ,\tag{54a}$$

$$q=(Q_n-Q_{n\,low})/(Q_{n\,up}-Q_{n\,low}) .\tag{54b}$$

According to Eq.(**50**) for different clays *p* and *q* are in the ranges

$$0 \leq p \leq 1 \quad \text{and} \quad 0 \leq q \leq 1 \quad . \tag{55}$$

In the force of the $Q_n(P_n)$ dependence *p* and *q* from Eq.(**54**) are also interconnected, $q=q(p)$. With that $0 \leq q \leq 1$ increases when $0 \leq p \leq 1$ decreases (Fig.(**6a**)). Accounting for Eqs.(**50**) and (**54**) the conditions from Eqs.(**51**) and (**53**) mean (Fig.(**6a**))

$$q(0)=1 \ , \quad q'(0)=0 \ , \quad \text{and} \quad q''(0)=0 \ . \tag{56}$$

Using Eq.(**56**) one can present $q(p)$ at $p \ll 1$ (see Fig.(**6a**)) as

$$q(p)=1-Dp^3 \ , \quad 0 \leq p \ll 1 \tag{57}$$

where *D* is some constant that should be found. Accounting for Eq.(**55**) we assume that the mutually opposite dependences $q(p)$ ($0 \leq p \leq 1$) and $p(q)$ ($0 \leq q \leq 1$) are symmetrical, that is, have the same mathematical form (that will be justified by the available data in Sections **5** and **6**). Then, based on the symmetry and Eq.(**57**) one can write (see Fig.(**6a**))

$$p(q)=1-Dq^3 \ , \quad 0 \leq q \ll 1 \ . \tag{58}$$

Note that the *D* coefficient in Eqs.(**57**) and (**58**) is the same in the force of the symmetry. We rewrite Eq.(**57**) in the form

$$q(p)=(1-p^3)^D \ , \quad 0 \leq p \ll 1 \ . \tag{59}$$

That is equivalent to Eq.(**57**) at $0 \leq p \ll 1$, but can be used as an approximation of $q(p)$ in the wider range of *p*, adjacent to *p*=0. Using similar considerations we also rewrite Eq.(**58**) in the form

$$p(q)=(1-q^3)^D \ , \quad 0 \leq q \ll 1 \ . \tag{60}$$

From Eq.(**60**) the $q(p)$ dependence at *p* close to unity can be written as

$$q(p)=(1-p^{1/D})^{1/3} \ , \quad 1-p \ll 1 \ . \tag{61}$$

Note that Eq.(**61**) (similar to Eq.(**59**)) can also be used as an approximation of $q(p)$ in the wider range of *p*, adjacent to *p*=1. We should "sew" $q(p)$ from Eqs.(**59**) and (**61**) in a point $p=p_o$ (see Fig.(**6a**)). The *D* and $p_o$ values are found from the smoothness conditions of $q(p)$ at $p=p_o$ as

$$q(p_o-)=q(p_o+) \quad \text{and} \quad q'(p_o-)=q'(p_o+) \tag{62}$$

($p_o-$ and $p_o+$ correspond to approaching to $p=p_o$ from the left and right, respectively). As a result we obtain $p_o \cong 0.795$ and $D \cong 0.3286$. As it should be, according to the above symmetry of $q(p)$ and $p(q)$ these $p_o$ and *D* values fulfill the equations (cf. Eqs.(**59**) and (**61**); see Fig.(**6a**))

$$p_o=(1-p_o^3)^D \quad \text{or} \quad p_o=(1-p_o^{1/D})^{1/3} \ . \tag{63}$$

Finally, the $q(p)$ dependence is written as (see Fig.(**6a**))

$$q = \begin{cases} (1-p^3)^{0.3286}, & 0 \leq p \leq 0.795 \\ (1-p^{1/0.3286})^{1/3}, & 0.795 \leq p < 1 \end{cases} . \tag{64}$$

In practical calculations one can also use the simple dependence as

$$q(p)=(1-p^3)^{1/3} \ , \quad 0 \leq p \leq 1 \ , \tag{65}$$

remembering that Eq.(**64**) is still more substantiated.

Using $q(p)$ from Eq.(65) and $q \leftrightarrow Q$, $p \leftrightarrow P$ relations (Eq.(54)) one can write $Q_n(P_n)$ as (see Fig.(6b))

$$Q_n(P_n) = Q_{n\ low} + (Q_{n\ up} - Q_{n\ low}) \cdot q((P_n - P_{n\ low})/(P_{n\ up} - P_{n\ low})) \ . \tag{66}$$

To estimate $Q_n$ one can take $Q_{n\ up}$, $Q_{n\ low}$, $P_{n\ up}$, and $P_{n\ low}$ from Eq.(50). However, more accurate (universal) presentation for $Q_n(P_n)$ is obtained by estimating these four values from positions of two extreme points in the ($P_n$, $Q_n$) plane (Fig.(6b); see Sections **5** and **6**). The experimental checking of Eq.(66) is considered in Sections **5** and **6**.

Thus, results of this and the previous section enable one to estimate the clay $Q$ factor as a universal function of the relative water content, $\zeta$ based on the clay parameters $v_s$ and $v_z$.

### 3.12. Input Physical Parameters of the Two-Factor Clay Water Retention Model

The $Q(\zeta)$ (Sections **3.10-3.11**) and $H(\zeta)$ (Sections **3.3-3.9**) factors give the clay water retention curve from the totally modified two-factor model, $h(\zeta)=H(\zeta)Q(\zeta)$ where $\zeta = \overline{w}/\overline{w}_M$. The modified $h(\overline{w})$ curve is determined by the same input physical parameters of the clay as in the basic model [22], $v_s$, $v_z$, $r_{mM}$, and the density of clay solids $\rho_s$. Note that $r_{mM}$ can be estimated through $v_z$ as follows. The maximum external size of 3D pores (at $\overline{w} = \overline{w}_M$), $r_{mM}$, playing the part of a characteristic scale, is connected with the maximum size of clay particles in oven-dried state, $r_{mz}$ as $r_{mM} = r_{mz} v_z^{-1/3}$ [25]. If we take $r_{mz} \cong 2\mu m$ (according to the generally accepted definition of the maximum size of clay particles in the oven-dried state) $r_{mM}$ is estimated to be $r_{mM} \cong 2v_z^{-1/3}$ (μm). This result is used in Section **5.2**.

## 4. EFFECT OF INTRA-AGGREGATE STRUCTURE ON SOIL WATER RETENTION CURVE

Now, to transit from the contributive-clay water retention to that of the clay-containing soil we should consider the transformation of the contributive-clay water content ($\overline{w}$) to that of the soil ($W$) (see Section **2**) following the recent model of the reference shrinkage curve [16, 18, 19]. This model relies on the new concepts of an intra-aggregate soil structure (Fig.(1)): (a) the existence and dewatering of a deformable, but non-shrinking superficial aggregate layer (interface layer) at any clay content as well as (b) the existence and volume increase of intra-aggregate lacunar pores at clay content lower than the critical one. We only touch on a point of the model that is relevant to finding the $W=W(\overline{w})$ relation between water contents of contributive clay ($\overline{w}$) and the soil as a whole ($W$). According to Eq.(3) there are two contributions to the water content of a soil as a whole, $W$: the contribution of the interface layer ($\omega$) and that of intra-aggregate matrix ($w'$) (Fig.(1)). According to Eq.(7) $w' = c\overline{w}/K$. To find the $W=W(\overline{w})$ relation we need dependence $\omega = \omega(w')$. One can write the water contribution of the interface layer (Fig.(1)), $\omega(w')$ [16] as

$$\omega(w') = \begin{cases} 0, & 0 \leq w' < w'_s \\ \rho_w U_i \Pi F_i(w'), & w'_s \leq w' < w'_h, \end{cases} \tag{67}$$

where $\rho_w$ is the water density; $U_i$ is a (constant) contribution of the interface layer (Fig.(1)) to the specific volume of the soil; $\Pi$ is a (non-shrinking) clay porosity of the interface layer (Fig.(1)) ($\Pi$ coincides with the soil porosity stipulated by clay matrix pores in the maximum swelling state); $F_i(w')$ is the volume fraction of the water-filled (non-shrinking) clay pores of the interface layer at a given $w'$ value; $w'_s$ corresponds to the end point of the structural shrinkage; $w'_h$ ($=W_h/K$) corresponds to the maximum swelling point. The $F_i(w')$ dependence exists in two variants, each of which is defined by the pore-size distribution of the interface and intra-aggregate clays (Fig.(1)). One can find details of the $\omega(w')$ calculation in [16,18], and the qualitative view of $\omega(w')$ in [17,19]. Input parameters for the $\omega(w')$ calculation (Eq.(67)) and correspondingly for the $W(\overline{w})$ calculation (Eq.(7)) of a soil include: the density of solids ($\rho_s$), relative volume of contributive-clay solids ($v_s$), relative volume of contributive clay in the oven-dried state ($v_z$), soil clay content ($c$), aggregate/intra-aggregate mass ratio ($K$), and specific volume of lacunar pores in the aggregates at maximum swelling ($U_{lph}$). There are different ways to calculate $U_i$ and $\Pi$ (see Eq.(67)). For instance, one finds subsequently the relative solid volume, $u_s$ of the soil to be

$$u_s = v_s[c + v_s(1-c)]^{-1}, \tag{68}$$

the maximum swelling water content of the soil, $W_h$ to be

$W_h = 0.5(1/u_s - 1)(\rho_w/\rho_s)$, (69)

the specific solid volume as $1/\rho_s$, and the specific aggregate volume, $U_h$ at $W=W_h$ to be

$U_h = 1/\rho_s + W_h/\rho_w + U_{lph}$. (70)

Finally,

$U_i = U_h(1 - 1/K)$ (71)

and

$\Pi = 1 - (1/\rho_s + U_{lph})/U_h$. (72)

For the calculation of $w'_s$ and $F_i$ (see Eq.(67)) at given $\Pi$ and clay shrinkage curve, $v(\zeta)$ [25, 26] see Eqs.(20)-(29) from [16].

The set of the necessary parameters ($\rho_s$, $v_s$, $v_z$, $c$, $K$, and $U_{lph}$) for the calculation of $W(\overline{w})$ includes those that are needed for the calculation of $h(\overline{w})$ (see Section **3.12**), $\rho_s$, $v_s$, and $v_z$ (if $r_{mM}$ is estimated through $v_z$). For data ($\rho_s$, $v_s$, $v_z$, $c$, $K$, and $U_{lph}$) accessibility see Section **5**.

Finally, note that in the above calculation of the aggregate surface layer contribution, $\omega(w')$ to the total water content $W$ (that follows [16]) we neglect the adsorbed water film that is present in the clay of the aggregate surface layer because its contribution to $W$ is small compared with that of the adsorbed water film that is in the clay of the intra-aggregate matrix (see Fig.(**1**); Section **3.8**). The reason for this is the small pore surface area of clay in the non-shrinking aggregate surface layer compared with that in the intra-aggregate matrix.

## 5. EXPERIMENTAL VALIDATION

### 5.1. Possible Ways to Estimate the Necessary Input Data

To substantiate the two-factor model of a soil water retention curve one needs data on the six above soil parameters to predict its water retention and independent data on the observed soil water retention to compare the prediction and observation (we imply the physical, but not curve-fitting comparison). In principle, the data on the indicated physical parameters can be obtained independently of the observed water retention and shrinkage curves of a soil. Indeed, $\rho_s$ is measured by standard methods [30]; $v_s$ and $v_z$ can be estimated from the oven-dried specific volume of the clay and water content at maximum swelling of the clay [16]; clay content, $c$ is measured by standard methods [31]; estimating the $K$ ratio by the maximum size of aggregates in the oven-dried state and the mean size of the soil solids (by their weight fractions) was recently suggested [20]; and finally, $U_{lph} = (W^*_h - W_h)/\rho_w$ (see Fig.(**7**)) where $W^*_h - W_h$ is a displacement between pseudo and true saturation lines. At the same time the available simultaneous data on the six indicated parameters, are missing because the overwhelming majority of corresponding works containing measured soil water retention curves are eventually oriented to curve-fitting when the above parameters are just not needed. As a result, although there are many works with data on experimental soil water retention curves, the latter are not accompanied with simultaneous data on $\rho_s$, $v_s$, $v_z$, $c$, $K$, and $U_{lph}$, and in most of the cases it is practically impossible to extract the necessary data from these works.

Nevertheless, in order to use available data we chose another way to estimate the necessary input data and took advantage of three following circumstances: (i) only the mutual independency between an observed water retention curve and data on the above six parameters is important; in all the other relations the origin of the parameter data is not essential for the aims of this work; (ii) the necessary data either accompany ($\rho_s$, $c$) the available observed water retention curves or can be extracted ($v_s$, $v_z$, $K$, $U_{lph}$) from a soil shrinkage curve (if it is also available) using the analysis that was recently described in detail [16,18]; (iii) single works are available where one simultaneously can find the experimental soil water retention and shrinkage curves for a number of soils.

### 5.2. Data Used

Based on the above three circumstances ((i)-(iii)) we considered six soils from Boivin et al. [12] (see Table **1**) who simultaneously presented experimental water retention and shrinkage curves. (We did not use data on two of the eight soils of [12], namely the soils from Figs.(**2c**) and (**5b**) of [12]. The reason is the non-simultaneous start of the soil shrinkage and growing of the soil suction in these figures that speak of some defect in shrinkage or suction data). The necessary data that enable us to predict the observed shrinkage curves for the soils are given in Table **1** (see explanations below). Since the experimental shrinkage curves in this work are only used as a source of the necessary parameter values (to predict the soil water retention curve), we only reproduce (in Fig.**(7)**) the data on the experimental shrinkage curve (white squares) for soil 4 (see Table **1**) from [12] as an example. In addition, Fig.**(7)** is used below for the illustration of different parameters. Data on $c$ and $\rho_s$ for soils 1 through 6 in Table **1** reproduce the data from [12].

To predict the reference shrinkage curve [18], one needs (see Table **1** and Fig.**(7)**): the oven-dried specific volume, $Y_z$; maximum swelling (gravimetric) water content, $W_h$; mean solid density, $\rho_s$, soil clay content, $c$; oven-dried structural porosity, $P_z$; the ratio of the aggregate solid mass to the solid mass of the intra-aggregate matrix, $K$; the lacunar factor, $k$; and water content $W_h^*$ with a displacement relative to $W_h$ that is similar to the displacement of the true saturated line relative to the pseudo one. If lacunar pores are absent at maximum swelling, $W_h^*=W_h$ (see Table **1**, soils 3 and 5). Additionally, if lacunar pores are absent at any water content in the course of shrinkage (as in Fig.**(1a)**), $k=0$ (see Table **1**, soil 5).

The $Y_z$, $W_h$, and $W_h^*$ values for indicated soils (see Fig.**(7)**) were estimated from the initial and final points of the corresponding shrinkage curves (Table **1**).

In estimating the structural porosity in the oven-dried state, $P_z$ (Table **1**) we took into account that $P_z$ differs from zero if the shrinkage curve has a horizontal section at water content $W>W_h$, that is, higher than the maximum swelling point [18]. The size of the section determines the specific volume of the structural (inter-aggregate) pores, $U_s$ and $P_z=U_s/Y_z$. If $U_s=0$ $P_z=0$ (as in Fig.**(7)**). Note that for soils 3 and 5 (Table **1**) $P_z>0$.

In this work $K$ was estimated from the experimental shrinkage curves (Table **1**) using its definition as the $W_h/w_h'$ ratio (see Fig.**(7)**). For an independent way to estimate $K$ see [20].

The soil lacunar factor $k$ by definition, is a micro-parameter of the intra-aggregate structure (Fig.**(1)**) that determines at $c<c^*$ the fraction of the clay matrix pore volume decrease that is transformed to the lacunar pore volume increase inside the aggregates (at $c>c^*$ $k=0$) [18,19]. Here, in estimating $k$ (Table **(1)**) we used the following important result: the $k$ micro-parameter is simply connected with such immediately observed macro-parameter of soil shrinkage as the slope $S$ of the reference shrinkage curve in the basic shrinkage area (Fig.**(7)**, $W_n<W<W_s$), $k=1-S\rho_w$.

Parameters $v_s$, $v_z$ (Table **(2)**) were also estimated for soils 1 through 6 in the course of the construction of the reference shrinkage curve for each soil (see as an example Fig.**(7)**) according to the approach from [16,18].

Thus, parameters $\rho_s$, $v_s$, $v_z$, $c$, $K$, $U_{lph}$ (or $W_h$ and $W_h^*$) from Tables **1** and **2** were used to predict the water retention curves of soils 1 through 6 based on the analysis from Sections **2-4**.

For example, Figs.**(8)** and **(9)** reproduce the data for the comparison (white circles), on the experimental water retention curves of soils 4 and 5 (Tables **1** and **2**) from Figs.**(2e)** and **(2f)** of [12]. Note that the data from [12] only cover the $W_n<W\leq W_h$ range of water content ($W_n$ is the end point of basic shrinkage).

## 5.3. Data Analysis

First, we checked the possibility of the clay $Q$ factor presentation as $Q(\zeta,Q_n)$ (Section **3.10**) at some definite $Q_n$ value. With this aim in mind, we calculated the water retention curve, $h(W)$ for each used soil as was described in Sections **2-4** using data on $\rho_s$, $v_s$, $v_z$, $c$, $K$, and $U_{lph}$ from Tables **1** and **2** and some $Q_n$ value from the range $Q_{nmin}<Q_n<Q_{nmax}$ ($Q_{nmin}$ and $Q_{nmax}$ for the soil from Eq.**(49)**). The possibility of the clay $Q$ factor presentation as $Q(\zeta,Q_n)$ was checked by the fitting of the predicted $h(W)$ (see Figs.**(8)** and **(9)**, solid lines for soils 4 and 5 as an example) to the experimental data (see Figs.**(8)** and **(9)**, white circles for soils 4 and 5 as an example) in the course of the $Q_n$ variation (between $Q_{nmin}$ and $Q_{nmax}$) for each soil (at given $\rho_s$, $v_s$, $v_z$, $c$, $K$, and $U_{lph}$) and using the least-square criterion. To check the feasibility of the $Q(\zeta,Q_n)$ presentation for each soil we estimated the goodness of fit, $r^2$, the best-fitted $Q_n$ value, and standard error, $\delta h$ of the experimental points $h_{ex}(W_i)$ (i=1,…, N) ($\delta h$ was calculated according to [32] simultaneously with $r^2$ and $Q_n$). Table **3** shows the $r^2$, $Q_n$, and $\delta h$ values that were found. See in Section **6** the results of the data analysis.

Second, we checked the universal dependence $Q_n(P_n)$ from Section **3.11** (Eq.**(66)**). Preliminarily, to construct this dependence we estimated $P_{n\,low}$, $P_{n\,up}$, $Q_{n\,low}$, and $Q_{n\,up}$ values entering Eq.**(66)**. With this aim in mind, we first found the experimental points $(P_{nj}, Q_{nj})$ (in the $P_n,Q_n$ plane; Fig.**(6b)**) for the six clays (j=1,…, 6) contributing to the six soils that were considered at $h(W)$ fitting (see above). With that, $Q_{nj}$ are the corresponding fitted $Q_n$ values from Table **3** and $P_{nj}=1-v_{sj}/v_{nj}$ are estimated for the clays (at $v_s$ and $v_z$ given for each clay in Table **2**) through the calculation of $v_n=v_s+(1-v_s)\zeta_n$ and $\zeta_n$ by $v_s$ and $v_z$ [25,26]. The $P_{nj}$ values (as well as $\zeta_z$, $\zeta_n$, and $v_n$) are also indicated in Table **3**. Figure **(6b)** shows the points $(P_{nj}, Q_{nj})$ (white squares with numbers j=1,…, 6). Then, to construct the theoretical dependence $Q_n(P_n)$ (Eq.**(66)**) we defined $P_{n\,low}$ and $Q_{n\,up}$ as $P_n$ and $Q_n$ for soil 3 (see Table **3** and point 3 in Fig.**(6b)**) as well as $P_{n\,up}$ and $Q_{n\,low}$ as $P_n$ and $Q_n$ for soil 2 (see Table **3** and point 2 in Fig.**(6b)**). Soils 2 and 3 were taken based on the obvious considerations connected with the extreme positions of the corresponding points in Fig.**(6b)**. The solid curve in Fig.**(6b)** shows the universal dependence $Q_n(P_n)$ (Eq.**(66)**) with indicated $P_{n\,low}$, $P_{n\,up}$, $Q_{n\,low}$, and $Q_{n\,up}$ and $q(p)$ from Eq.**(64)** (Fig.**(6a)**). To check $Q_n(P_n)$ dependence one should compare the relative positions of the ($Q_n$, $P_n$) points corresponding to soils 1, 4, 5, and 6 [that did not participate in the construction of $Q_n(P_n)$] and the solid curve in Fig.**(6b)**. See in Section **6** the results of the analysis.

Finally, to check the two-factor model as a whole we should compare, for each of soils 1, 4, 5, 6, the predicted soil water retention curves $h(W, Q_n)$ with the fitted $Q_n$ value (see Table **3**; Figs.**(8)** and **(9)**, solid lines for soils 4 and 5 as an example) and the theoretical $Q_n(P_n)$ value that follows from Eq.**(66)** and lies on the solid curve in Fig.**(6b)** at a given $P_n$ (that corresponds to soils 1, 4, 5, and 6, see Table **3**). See in Section **6** the results of the analysis.

# 6. RESULTS AND DISCUSSION

### 6.1. About the $Q$ Factor Presentation through the $Q_n$ Parameter

In Figs.**(8)** and **(9)** (for soils 4 and 5 as an example) discrepancies between the experimental points and the $h(W)$ curve (solid one) that was fitted using the $Q(\zeta, Q_n)$ presentation of the $Q$ factor (Eqs.**(43)**, **(45)-(47)**), do not surpass approximately two standard errors, $\delta h$ from Table **3** (the same relates to soils 1-3 and 6) These estimates along with the high $r^2$ values (Table **3**) speak in favor of the feasibility of $Q$ factor presentation from Eqs.**(43)**, **(45)-(47)**.

### 6.2. About the Theoretical Dependence of $Q_n(P_n)$

Accounting for the errors $\delta h$ (Table **3**) of the experimental $h_{ex}(W)$ points (white circles in Figs.**(8)** and **(9)**), two-factor model approximations (Section **3**), and computational errors, one can estimate the errors, $D_Q$ of the fitted $Q_n$ values (Table **3**) as $D_Q<0.05$ by order of magnitude. On the other side, as one can see, in Fig.**(6b)** the discrepancy, $|Q_n-Q_{ntheor}|$ between the $Q_n$ values of the six points and corresponding $Q_{ntheor}$ values of solid line, $Q_n(P_n)$ (Eq.**(66)**) does not surpass ~0.05. That is, $|Q_n-Q_{ntheor}|\leq D_Q$. Therefore, the arrangement of the ($P_{nj}$, $Q_{nj}$) points (j=1,…, 6) in the ($P_n$, $Q_n$) plane (Fig.**(6b)**) does not contradict the theoretical dependence $Q_n(P_n)$ (Eq.**(66)**).

### 6.3. About the Two-Factor Model as a Whole

Because for the six soils $|Q_n-Q_{ntheor}|\leq D_Q$ [$Q_{ntheor}=Q_n(P_n)$] the predicted $h(W, Q_{ntheor})$ practically coincides with the fitted $h(W, Q_n)$ for each soil (see the solid line in Figs.**(8)** and **(9)** for soils 4 and 5 as an example). Therefore the discrepancies between the experimental $h_{ex}(W)$ points (white circles in Figs.**(8)** and **(9)**) and theoretically predicted soil water retention curve, $h(W, Q_{ntheor})$ (solid line in Figs.**(8)** and **(9)**) do not surpass two standard errors, $\delta h$ from Table **3**. This result shows that the soil water retention curves predicted by the two-factor model (Sections 2-4) for the six soils from [12] are in the good agreement with the corresponding experimental curves from [12].

It should be noted that the water content range of the predicted suction head ($W_a \leq W \leq W_h$) is wider than that of the observed suction head in [12] ($W_n < W \leq W_h$) (see as an example Figs.**(8a)** and **(9a)**). Although the theoretical $h(W)$ dependences in the different parts of the water content range are closely connected with each other (according to Sections **2-4**), and thereby the experimental confirmation for the $W_n < W \leq W_h$ range is in part the confirmation for the $W_a \leq W \leq W_n$ range also, nevertheless, the comparison of the predicted $h(W)$ and independent data for the $W_a \leq W \leq W_n$ range is desirable in the future.

Several additional points should be mentioned:

(a) Table **4** contains the estimates of all the values connected with the boundary water content $\zeta_a$ (see Fig.**(5)**). These results show the characteristic order of magnitudes. Table **4** shows that among the six considered soils only in the case of soil 3 does $\zeta_a=\zeta_z$ ($\zeta_z$ see in Table **3**; or $W_a=W_z$ in Table **2**), $\rho'_{min}(\zeta_a)>2l_a$, and $F_a=F_z$; for the other soils $\zeta_a<\zeta_z$, $\rho'_{min}(\zeta_a)=2l_a$, and $F_a<F_z$ (cf. Section **3.8**). Note that the maximum value $L_a$ (Table **4**) of the summary perimeter of clay pore tubes (per unit surface area of their cross-section) containing only the maximum adsorbed water film, varies in a small range 6.2-7.9 μm$^{-1}$ for the soils under consideration. According to Table **4** the ratios $\rho'_{min}(\zeta_a)/r_{mM}$ and $\rho'_m(\zeta_a)/r_{mM}$ (see Fig.**(5)**) also vary in a small ranges. The values $\rho'_{min}(\zeta_a)$ and $\rho'_m(\zeta_a)$ (Table **4**) characterize the total range of the size variation of the internal pore-tube cross-section in the contributive-clay matrices at $\zeta=\zeta_a$. Finally, it is worth noting that the maximum thickness of the adsorbed water film, $l_a$ is in the range 4 10$^{-2}$-10$^{-1}$μm (Table **4**) that is in agreement with the estimates [33] flowing out of the general physical considerations (10$^{-3}$-10$^{-1}$μm).

(b) As one can see in Table **2** for all the six considered soils ($W_n-W'$)<<$W_n$ (cf. Section **3.9**).

(c) At a given set of the input parameters ($\rho_s$, $v_s$, $v_z$, $c$, $K$, and $U_{lph}$) the soil shrinkage curve has two possible variants in the vicinity of the maximum swelling point, $W_h$ [16,18,19]. In the prediction of the soil water retention curve $h(W)$ we tried both the possible $W(\overline{w})$ dependences. The experimental data [12] on the retention curves are in agreement with the predicted $h(W)$ only for the $W(\overline{w})$ variant leading to the shrinkage curve that is convex upward in the vicinity of the $W=W_h$ point (see [16]). Figures **(8)** and **(9)** show the results that relate namely to this variant.

(d) The specification of the universal dependence $Q_n(P_n)$ (the solid curve in Fig.**(6b)**) is possible as data accumulation with other soils. However, as judged by Fig.**(6b)** it should be very small because extreme points 3 and 2 in Fig.**(6b)** are on the horizontal and vertical sections of the $Q_n(P_n)$ line, respectively.

(e) About additional results of this work. This work is aimed at the physical modeling of clay soil water retention based on the generalization of the two-factor model of clay water retention [22] as well as on the new concepts of intra-aggregate soil structure (Fig.**(1)**) [16,18,19]. Hence, the above results relative to the soil water retention give the additional validation of both the two-factor model (in addition to the results relating to pure clay from [22]) and the new concepts (in addition to the results of soil shrinkage prediction from [16,18,19]).

(f) About the case of sufficiently small clay content. In this case the differences between intra-aggregate lacunar pores (Fig.**(1)**) and inter-aggregate ones disappear because the aggregates themselves diminish and are reduced to separate sand and silt grains (see [18,20]). The above general approach to the soil water retention remains applicable in the case of the small clay content. However the contribution to the soil suction, stipulated by the intra-aggregate clay (Fig.**(2)**), is degenerated and the contribution of the "tail" (Fig.**(2)**) becomes the major one. For this reason the contributing sand for such soils should be considered as a "zero approximation", but not clay, as in this work.

Finally, it is worth reiterating an essential difference between the available models and the one presented in this work. These models are not physical ones in the exact sense of the word because they do not give the quantitative soil water retention prediction from a number of physical parameters. Even the models that start from (different for each model) fundamental concepts (e.g., [7,8,14]), are eventually reduced to curve-fitting. A number of input model parameters that they use can only be found by fitting to an observed water retention curve. Unlike that, the presented model only uses the physical input soil parameters that can be measured or estimated independently of an observed water retention curve. This difference is of principle importance for understanding the physical interrelations between soil structure and hydraulic functions, even though one can quickly measure a local soil water retention curve and even though there are good fitting approximations for the local curve.

## 7. CONCLUSION

The objective of this work is to derive the soil water retention from soil structure without curve-fitting and using only parameters with clear physical meaning that can be measured independently of an experimental retention curve. Two obvious key points underlying the work are: (i) the soil suction at drying ($h$) coincides with that of the soil intra-aggregate matrix and contributive clay; and (ii) both the soil suction and volume shrinkage at dewatering (drying) depends on the same soil water content ($W$). Together with some recent results these two simple points open the way to the physical prediction of soil water retention. Indeed, the two-factor (capillarity and shrinkage) model of clay water retention [22] enables one to connect clay suction ($h$) and clay water content ($\overline{w}$) based on the clay matrix structure [24,25] and without curve-fitting. Accounting for this result the key point (i) means that the same dependence ($h(\overline{w})$) takes place between soil suction ($h$) and water content of the contributive clay ($\overline{w}$). In addition, the reference shrinkage curve model [16,18,19], based on new concepts of intra-aggregate soil structure, permits one to connect the soil water content ($W$) at volume shrinkage with the water content of the contributive clay ($\overline{w}$). Accounting for this result key point (ii) implies that the same connection ($W(\overline{w})$) is kept when the soil suction increases at drying and shrinkage. Thus, knowing the $h(\overline{w})$ dependence for contributive clay from [22] and $W(\overline{w})$ dependence from [16,18,19] one comes to the parametric presentation of the soil water retention curve, $h(W)$ (with $\overline{w}$ as parameter). The realization of this approach required essential modification and specification of the available two-factor model for clay [22] and, in particular, taking the effect of adsorbed water film into account (Sections **3-4**). As a result the model as a whole includes the following input parameters: the density of solids ($\rho_s$), relative volume of contributive-clay solids ($v_s$), relative volume of contributive-clay in the oven-dried state ($v_z$), soil clay content ($c$), aggregate/intra-aggregate mass ratio ($K$), and specific volume of lacunar pores in the aggregates at maximum swelling ($U_{lph}$).

The substantiation of the model is based on available data simultaneously on the experimental water retention and shrinkage curves for six aggregated clay soils from [12]. The analysis of the shrinkage curves using the approach from [16,18] gives the necessary input for the soil water retention prediction (how the input data is obtained is not essential for the aims of this work). The major result of this work is the quite reasonable and promising agreement between the predicted (without curve-fitting) and observed water retention curves of the aggregated shrink-swell soils.

## NOTATION

*A*     constant of clay microstructure, dimensionless
*c*     soil clay content, kg kg$^{-1}$
*D*     coefficient in Eqs.**(57)-(61)**, **(63)**, dimensionless

| | |
|---|---|
| $D_Q$ | estimate of standard error of $Q_n$, dimensionless |
| $F(\zeta)$ | saturation degree, dimensionless |
| $F_a$ | $F$ value at $\zeta=\zeta_a$, dimensionless |
| $F_z$ | $F$ value at $\zeta=\zeta_z$, dimensionless |
| $G$ | coefficient in Eq.**(40)**, dimensionless |
| $g(\rho')$ | degree of water-filling the pore tubes of the internal $\rho'$ size, dimensionless |
| $H$ | one of two factors determining suction, kPa or cm of water |
| $h$ | soil (or clay, or intra-aggregate matrix) suction, kPa or cm of water |
| $h_o$ | suction at the maximum soil swelling (Fig.**(2)**), kPa or cm of water |
| $I(x)$ | function entering the pore-tube size distribution (Eq.**(17)**), dimensionless |
| $K$ | aggregate/intraaggregate mass ratio, dimensionless |
| $k$ | lacunar factor, dimensionless |
| $L(\zeta)$ | summary perimeter of pore tubes (per unit surface area of their cross-section) containing only the maximum adsorbed water film, µm$^{-1}$ |
| $L_a$ | $L$ value at $\zeta=\zeta_a$, µm$^{-1}$ |
| $l_a$ | maximum thickness of adsorbed water film, µm |
| $M(\zeta)$ | function determined after Eq.**(22)**, dimensionless |
| $P$ | clay matrix porosity, dimensionless |
| $P_n$ | clay porosity at the air-entry point, dimensionless |
| $P_{n\,low}$ | lower boundary of $P_n$ for different clays, dimensionless |
| $P_{n\,up}$ | upper boundary of $P_n$ for different clays, dimensionless |
| $P_z$ | oven-dried structural porosity of soil in oven-dried state, dimensionless |
| $p$ | relative value connected with $P_n$ (Eq.**(54a)**, dimensionless |
| $p_o$ | specific $p$ value of the $q(p)$ dependence, dimensionless |
| $Q$ | one of two factors determining suction, dimensionless |
| $Q_1,.,Q_3$ | coefficients in Eq.**(43)**, dimensionless |
| $Q_n$ | $Q(\zeta_n)$ value, dimensionless |
| $Q_{n\,low}$ | lower boundary of $Q_n$ for different clays, dimensionless |
| $Q_{nmax}$ | possible maximum value of $Q_n$ for a given clay, dimensionless |
| $Q_{nmin}$ | possible minimum value of $Q_n$ for a given clay, dimensionless |
| $Q_{n\,up}$ | upper boundary of $Q_n$ for different clays, dimensionless |
| $q$ | relative value connected with $Q_n$ (Eq.**(54b)**), dimensionless |
| $q_i(i=1,\ldots,6)$ | coefficients in Eq.**(42)**, dimensionless |
| $R(\zeta)$ | characteristic (internal) size of pore-tube cross-section of clay matrix, µm |
| $r_{mM}$ | maximum external size of clay pores at the liquid limit, µm |
| $r_0(\zeta)$ | minimum external size of clay pores, µm |
| $r^2$ | goodness of fit for the best-fitted $Q_n$ value, dimensionless |
| $U_{lph}$ | specific lacunar pore volume at maximum swelling, dm$^3$/kg |
| $U_s$ | specific volume of inter-aggregate pores, dm$^3$/kg |
| $v(\zeta)$ | relative clay volume, dimensionless |
| $v_h$ | relative clay volume at maximum swelling, dimensionless |
| $v_M$ | relative clay volume at liquid limit ($v_M=1$), dimensionless |
| $v_n$ | relative clay volume at air-entry point, dimensionless |
| $v_s$ | relative volume of clay solids, dimensionless |
| $v_z$ | $v$ value at the shrinkage limit of clay, dimensionless |
| $W$ | gravimetric soil water content, kg kg$^{-1}$ |
| $W_a$ | boundary of the exhausting capillary water, kg kg$^{-1}$ |
| $W_h$ | water content at maximum soil swelling, kg kg$^{-1}$ |
| $W_m$ | maximum soil water content, kg kg$^{-1}$ |
| $W_n$ | end point of soil basic shrinkage, kg kg$^{-1}$ |
| $W_s$ | end point of soil structural shrinkage, kg kg$^{-1}$ |
| $W_z$ | shrinkage limit, kg kg$^{-1}$ |
| $W'$ | "sewing" point (where $\rho'_f(\zeta)= \rho'_c(\zeta)$), kg kg$^{-1}$ |
| $w$ | water content of intra-aggregate matrix, kg kg$^{-1}$ |
| $w_h$ | $w$ value at the maximum swelling point, kg kg$^{-1}$ |
| $w_s$ | $w$ value at the end point of soil structural shrinkage, kg kg$^{-1}$ |
| $\bar{w}$ | water content of the contributive clay, kg kg$^{-1}$ |
| $\bar{w}_h$ | maximum swelling point of the clay, kg kg$^{-1}$ |
| $\overline{w}_M$ | clay liquid limit, kg kg$^{-1}$ |
| $\overline{w}_s$ | $\overline{w}$ value at the end point of soil structural shrinkage, kg kg$^{-1}$ |
| $w'$ | contribution of intra-aggregate matrix to soil water content, $W$, kg kg$^{-1}$ |
| $w'_h$ | $w'$ value at maximum swelling point, kg kg$^{-1}$ |
| $w'_s$ | $w'$ value at the end point of soil structural shrinkage, kg kg$^{-1}$ |
| $x(\rho')$ | argument of the $I(x)$ function in Eq.**(16)-(17)**, dimensionless |
| $Y_z$ | minimum specific volume of soil, dm$^3$ kg$^{-1}$ |
| $Z(\zeta)$ | function determined after Eq.**(22)**, dimensionless |
| | |
| $\alpha$ | constant of clay microstructure, dimensionless |
| $\alpha_c$ | contact angle, degrees |
| $\Gamma$ | surface tension of water, N/m |
| $\Delta$ | mean thickness of clay particles, µm |
| $\delta h$ | estimate of standard error of experimental suction values, kPa or cm of water |
| $\Delta W_m$ | maximum water content of capillary inter-aggregate pores, kg kg$^{-1}$ |
| $\delta(\zeta)$ | mean thickness of clay particle cross-sections, µm |
| $\zeta$ | relative water content of clay, dimensionless |
| $\zeta_a$ | relative water content corresponding to the maximum adsorbed film, dimensionless |
| $\zeta_h$ | relative water content of clay at maximum swelling, dimensionless |
| $\zeta_n$ | relative water content of clay at air-entry point, dimensionless |
| $\zeta_z$ | relative water content of clay at the shrinkage limit, dimensionless |
| $\zeta'$ | "sewing" point (Eq.**(40)**), dimensionless |
| $\rho$ | external pore-tube cross-section size, µm |
| $\rho_s$ | density of solids, kg dm$^{-3}$ |
| $\rho_w$ | water density, kg dm$^{-3}$ |
| $\rho_m(\zeta)$ | maximum external size of clay pore-tube cross-sections, µm |
| $\rho_o(\zeta)$ | minimum external size of clay pore-tube cross-sections, µm |
| $\rho'$ | internal pore-tube cross-section size of clay matrix, µm |
| $\rho'_m$ | maximum $\rho'$ value in clay matrix, µm |
| $\rho'_{min}$ | minimum $\rho'$ value in clay matrix, µm |
| $\rho'_c(\zeta)$ | maximum internal size of water-containing pore tubes, µm |
| $\rho'_f(\zeta)$ | maximum internal size of water-filled pore tubes, µm |
| $\varphi(\rho')$ | one-mode pore-tube cross-section size distribution of clay, dimensionless |
| $\omega$ | contribution of aggregate surface layer to soil water content, kg kg$^{-1}$ |
| $\omega_h$ | $\omega$ value at maximum swelling point, kg kg$^{-1}$ |

**Figure Captions**

**Fig.(1)**. Illustrative scheme of the internal structure of aggregates at a clay content: (a) $c>c^*$, without lacunar pores; and (b) $c<c^*$, with lacunar pores and possible non-totally contacting silt-sand grains where $c^*$ is the critical soil clay content (the modified Fig.(**3**) from [16]).

**Fig.(2)**. Illustrative graph of the soil water retention curve including a "tail" in the range $W>W_h$ stipulated by the capillary inter-aggregate porosity.

**Fig.(3)**. Illustrative graph showing the single-valued interrelations between water retention curves (drying branches) of clay ($h(\overline{w})$ - curve 1), intra-aggregate matrix ($h(w)$ - curve 2), and aggregated soil ($h(W)$ - curves 3 and 4; the latter correspond to two variants of the interface layer contribution to the soil water content, $\omega_1$ and $\omega_2$, respectively, [16,18]) as well as transitions between them. At a given suction $h=h(W)=h(w)=h(\overline{w})$ water contents $\overline{w}$, $w$, and $W$ are interconnected as: $w=c\overline{w}$ and $W=w/K+\omega(w/K)=c\overline{w}/K+\omega(c\overline{w}/K)$ (for $\omega(w')$ see [16,18]). The soil water content, $W_s$ and suction $h_s$ correspond to the end point of the structural shrinkage area of the soil.

**Fig.(4)**. General view of the $Q$ factor and relative $H$ factor of a clay. $\zeta_z$ is the shrinkage limit; $\zeta_h$ is the maximum swelling point; $\zeta_a$ is the water content that only corresponds to the adsorbed film of the maximum thickness; and $\zeta_n$ is the air-entry point. $Q_{nmin}$ and $Q_{nmax}$ determine the possible minimum and maximum of $Q_n$ for a given clay (see Section **3.10**).

**Fig.(5)**. Qualitative view of relative characteristic internal pore-tube cross-section sizes of a clay matrix against the relative water content (the modified Fig.(**4**) from [22]). "Relative" size means the ratio of a size to $r_{mM}$ (maximum pore size at liquid limit); subscript $i$ of $\rho'_i$ corresponds to the index of the shown curves, $i=1,\ldots,5$. 1-the maximum internal size of pore-tube cross-sections, $\rho'_m(\zeta)/r_{mM}$ at $0<\zeta<\zeta_n$: 2-the same size as on curve 1, but at $\zeta_n<\zeta<1$; 3-the maximum internal size of water-filled pore-tube cross-sections, $\rho'_f(\zeta)/r_{mM}$ at $\zeta_a<\zeta<\zeta_n$; 4-the maximum internal size of water-containing pore-tube cross-sections, $\rho'_c(\zeta)/r_{mM}$ at $\zeta_a<\zeta<\zeta_n$; 5-the minimum internal size of pore-tube cross-sections, $\rho'_{min}(\zeta)/r_{mM}$. The smooth curve composed of curve 2 at $\zeta_n<\zeta<\zeta_h$ and curve 4 at $\zeta_a<\zeta<\zeta_n$ gives the relative characteristic size, $R(\zeta)/r_{mM}$ that determines the $H$ factor as a function of the relative water content. $\zeta_a$, $\zeta_z$, $\zeta'$, $\zeta_n$, and $\zeta_h$ are relative water contents corresponding to the boundary of the exhausting capillary water, shrinkage limit, "sewing" point where $\rho'_f(\zeta')=\rho'_c(\zeta')$, air-entry point, and maximum swelling point, respectively. The black circle marks merging curves 3, 4, and 5 at $\zeta=\zeta_a$ (see Eq.(**20**)).

**Fig.(6a)**. The auxiliary function $q(p)$ (Eq.(**64**)) participating in the calculation of the $Q_n(P_n)$ dependence.

**Fig.(6b)**. The theoretical universal $Q_n(P_n)$ dependence (Eq.(**66**)) (solid line). Points 3 and 2 (corresponding to soils 3 and 2 in Tables **1-3**) participated in estimating the parameters of the dependence. The arrangement of other points visually show the agreement between the model (solid curve) and data used.

**Fig.(7)**. Shrinkage curve data (white squares) and prediction (solid line) corresponding to soil 4 in Tables **1-3**. The maximum relative difference $\delta=\max(|Y-Y_e|/Y_e)$ between the predicted ($Y$) and experimental ($Y_e$) values of the specific volume for the soil is equal to 0.004. The dotted line is parallel to the shrinkage curve in the basic shrinkage area. Dashed and dash-dot inclined lines are the true and pseudo saturation lines, respectively. The water contents $W_z$, $W_n$, $W_s$, $W_h$, and $W_h^*$, correspond to shrinkage limit, end-point of basic shrinkage, end-point of structural shrinkage, maximum swelling, and filling of lacunar pores (if they are filled in), respectively. The specific volumes $Y_z$ and $Y_h$ correspond to oven-dried state and maximum swelling, respectively. $w_h'$ is the maximum contribution of the intra-aggregate matrix to the total water content $W_h$ at maximum swelling. $U_{lph}=(W_h^*-W_h)/\rho_w$ is the specific volume of lacunar pores in the intra-aggregate matrix at maximum swelling.

**Fig.(8a)**. The water retention curve for soil 4 (see Tables **1-3**). The white circles present data from [12]. The solid line presents the model predicted $h(W)$ dependence in the total range of the water content for both the fitted $Q_n$ parameter and theoretically predicted, $Q_n(P_n)$ ($P_n$ is the porosity of the contributive clay at the air-entry point). Black circles on the curve correspond to the characteristic values of the soil water content with indicated marks (the sign and indices). These water contents respond to the boundary of the exhausting capillary water ($W_a$); shrinkage limit ($W_z$); "sewing" point (where $\rho'_f(\zeta')=\rho'_c(\zeta')$) ($W'$); end point of soil basic shrinkage ($W_n$); end point of soil structural shrinkage ($W_s$); and the maximum swelling point ($W_h$).

**Fig.(8b)**. The part of Fig.(**8a**) for the limited range of soil water content where there are the experimental data from [12].

**Fig.(9a)**. As in Fig.(**8a**) for soil 5.

**Fig.(9b)**. The part of Fig.(**9a**) for the limited range of soil water content where there are the experimental data from [12].

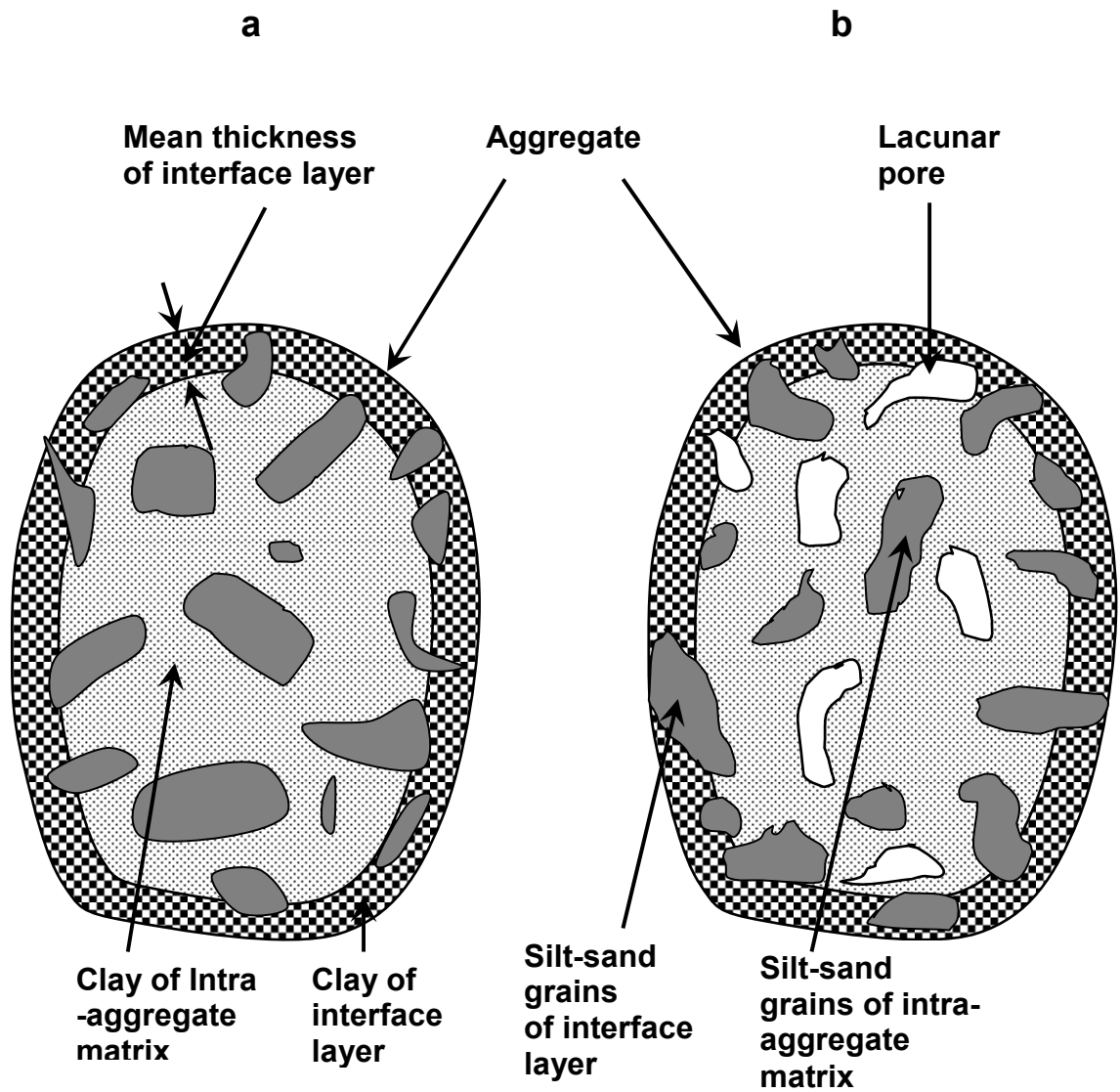

Fig.1

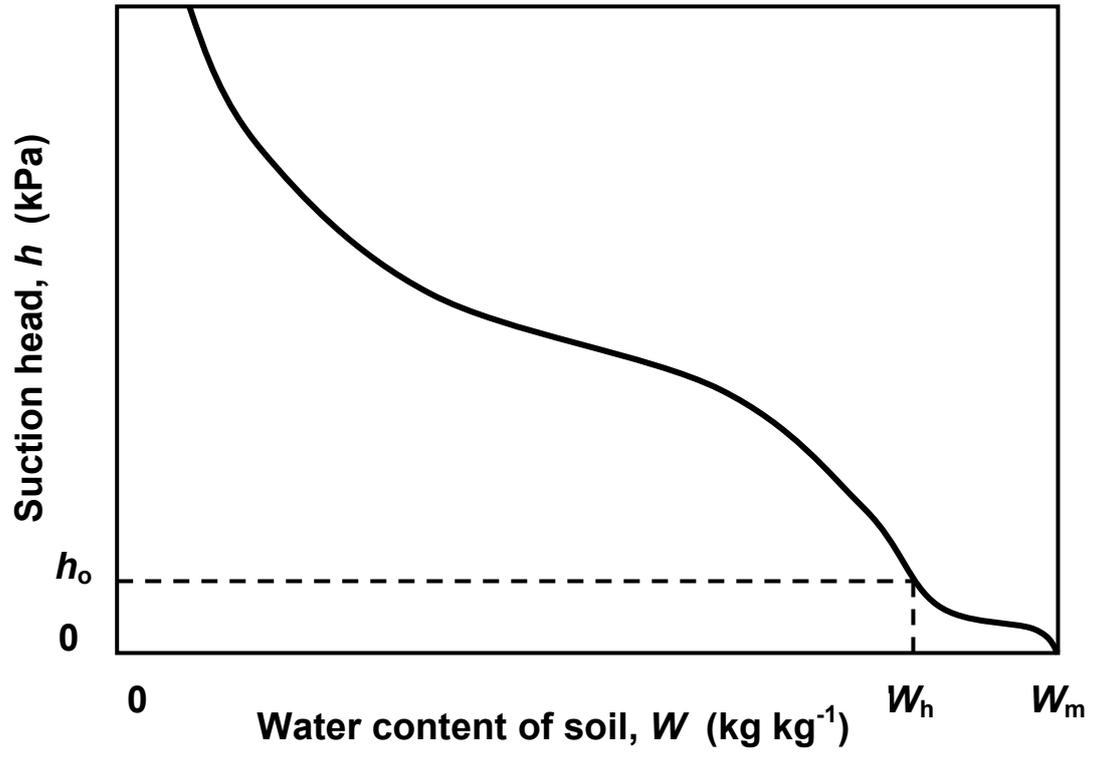

Fig.2

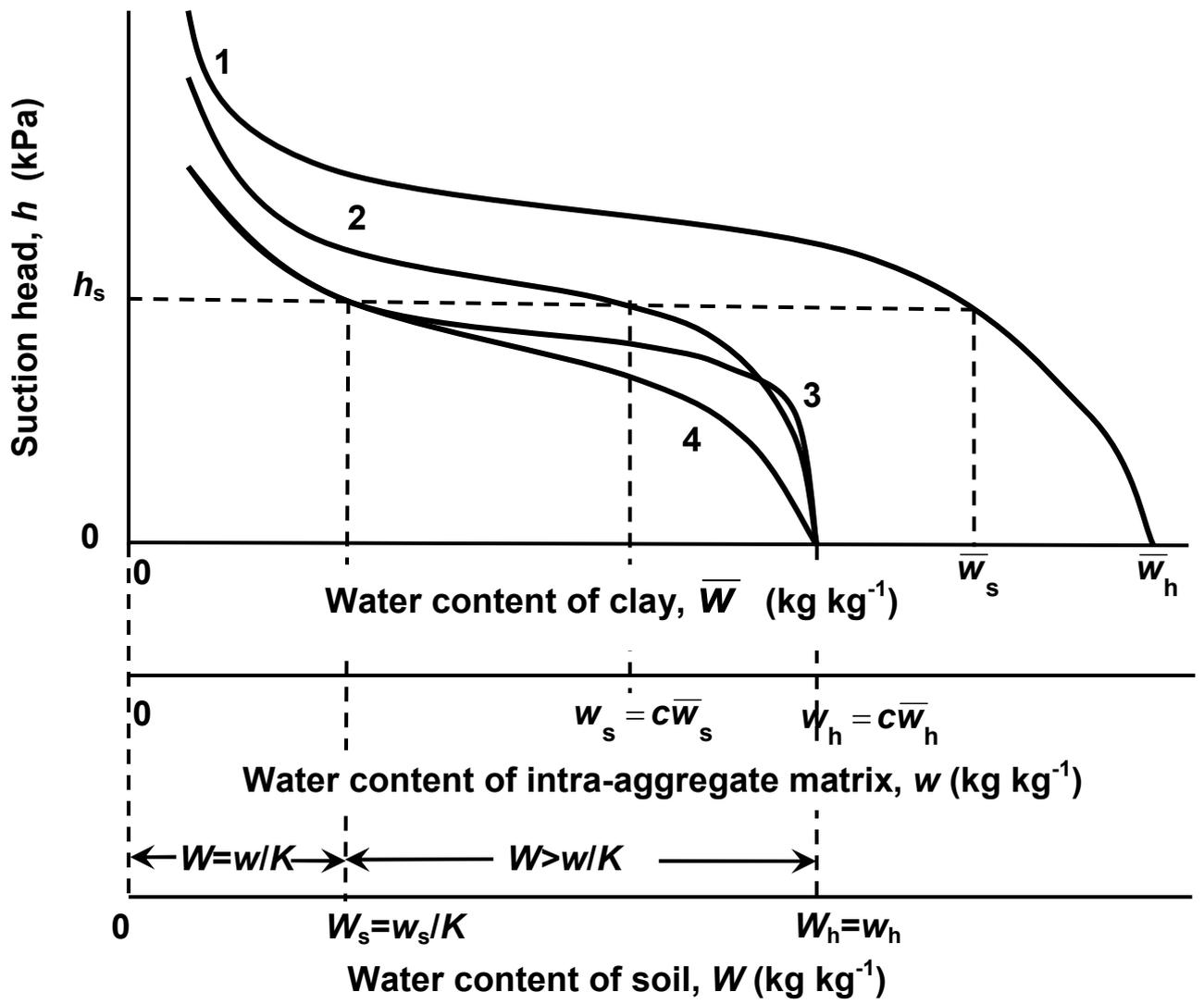

Fig.3

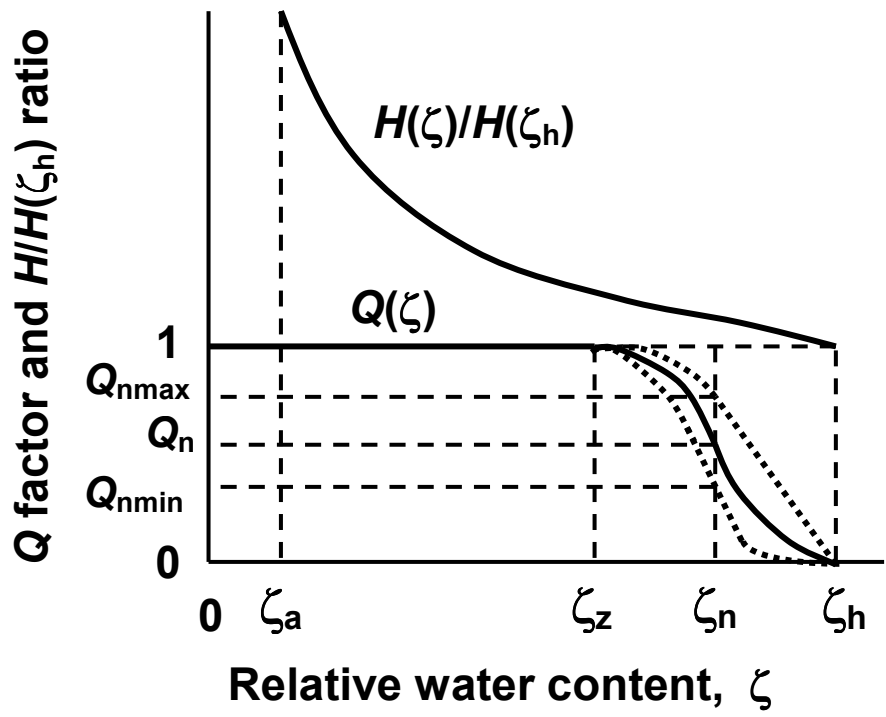

Fig.4

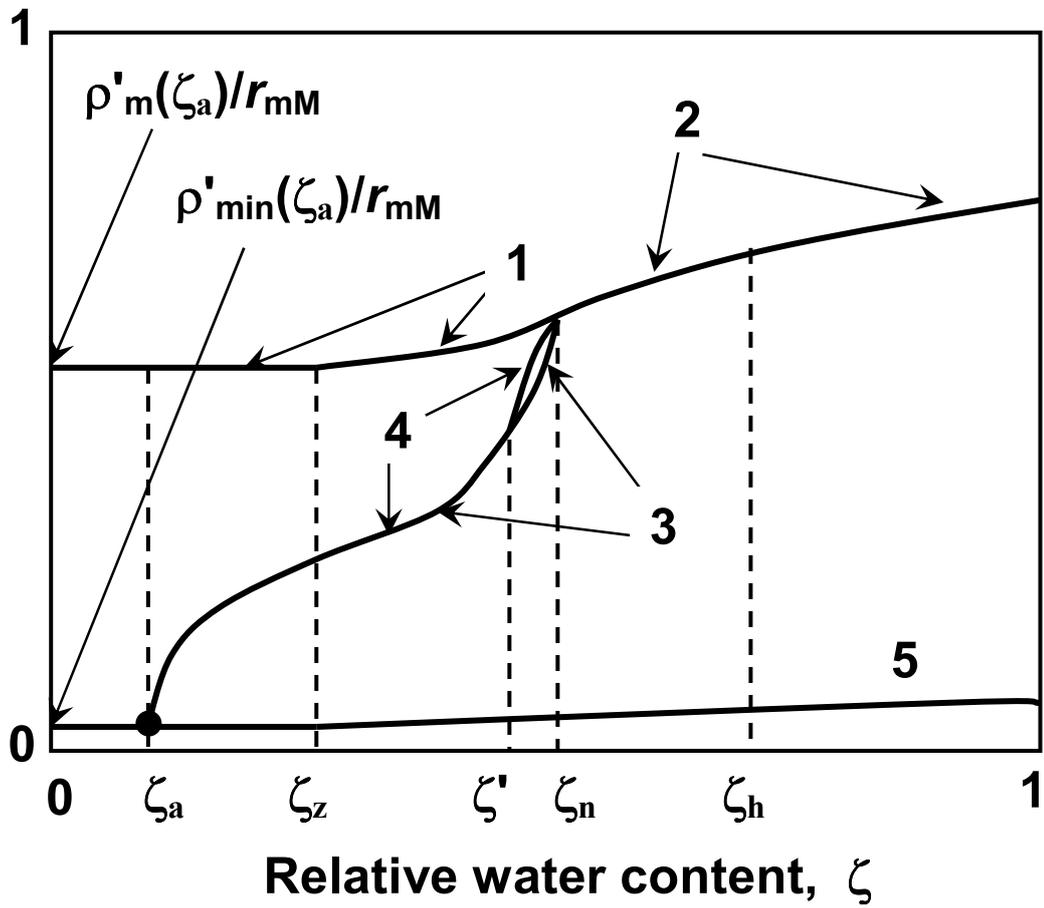

Fig.5

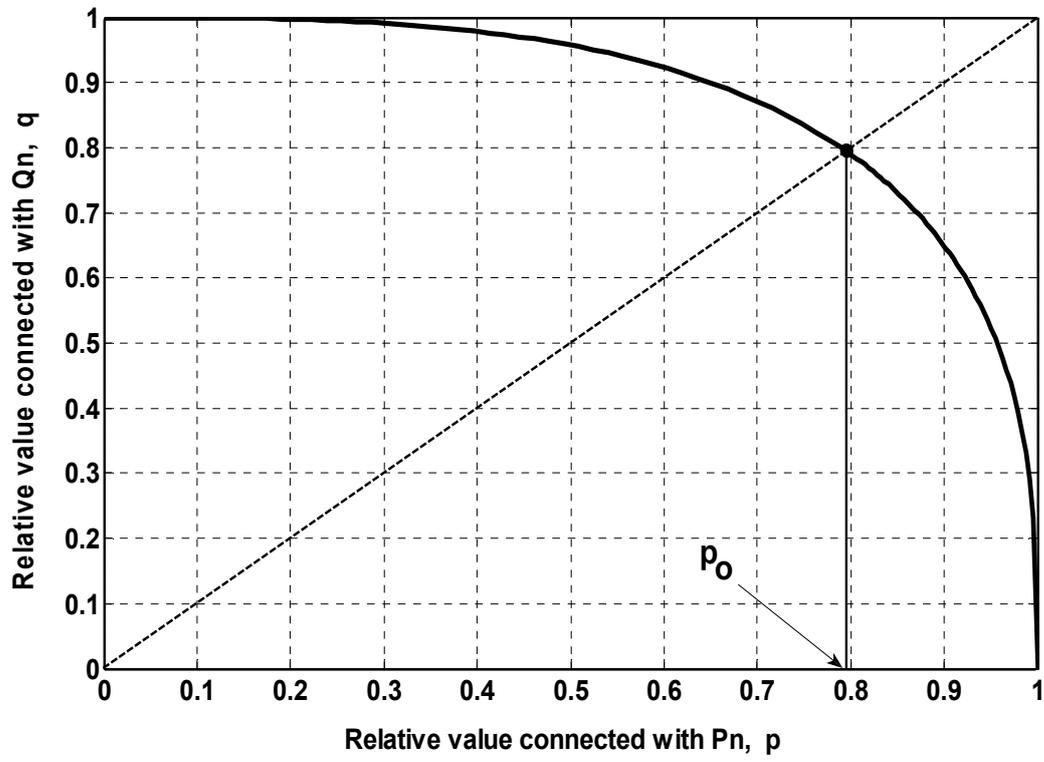

Fig.6a

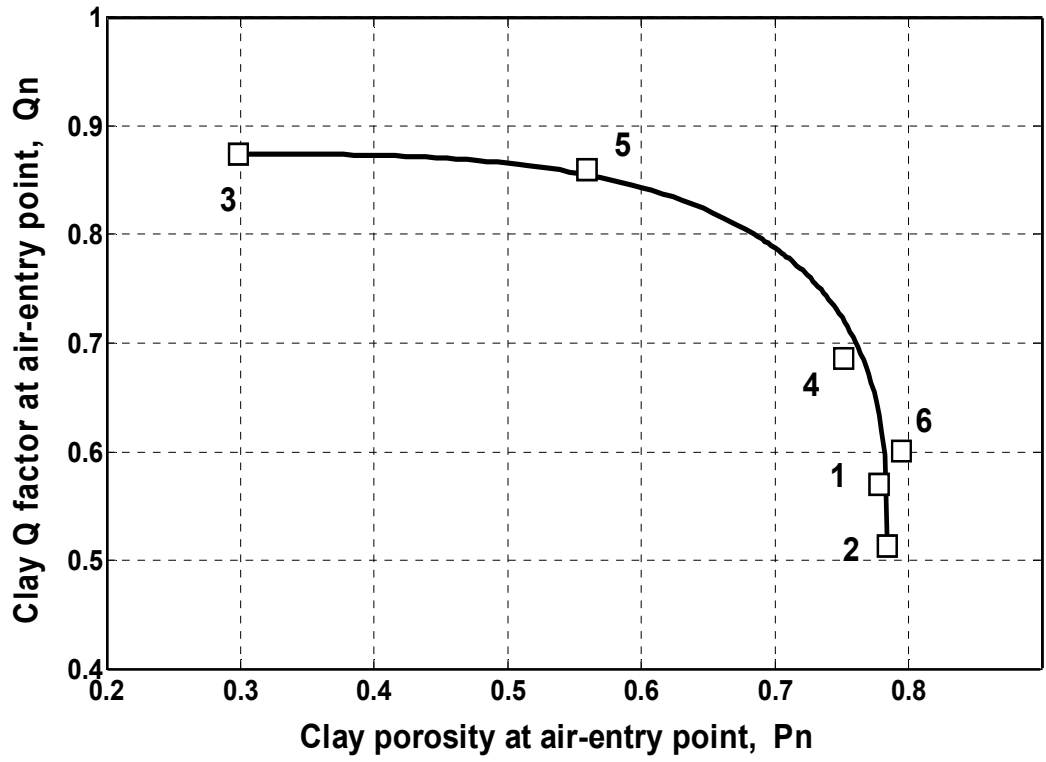

Fig.6b

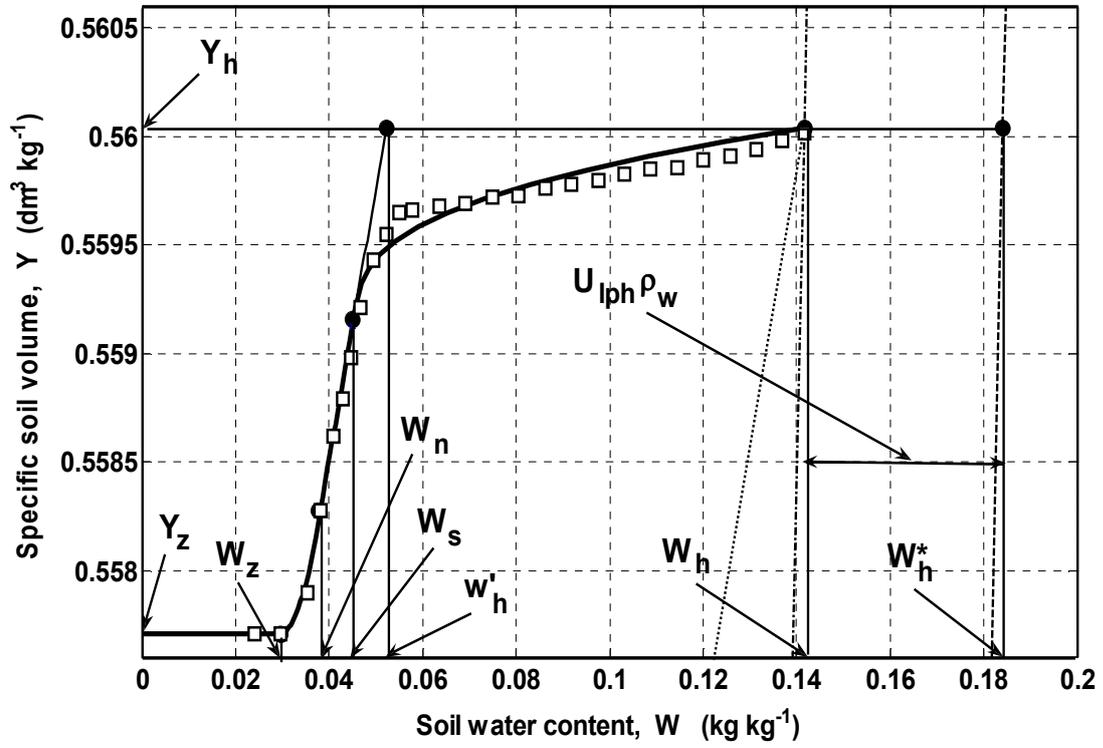

Fig.7

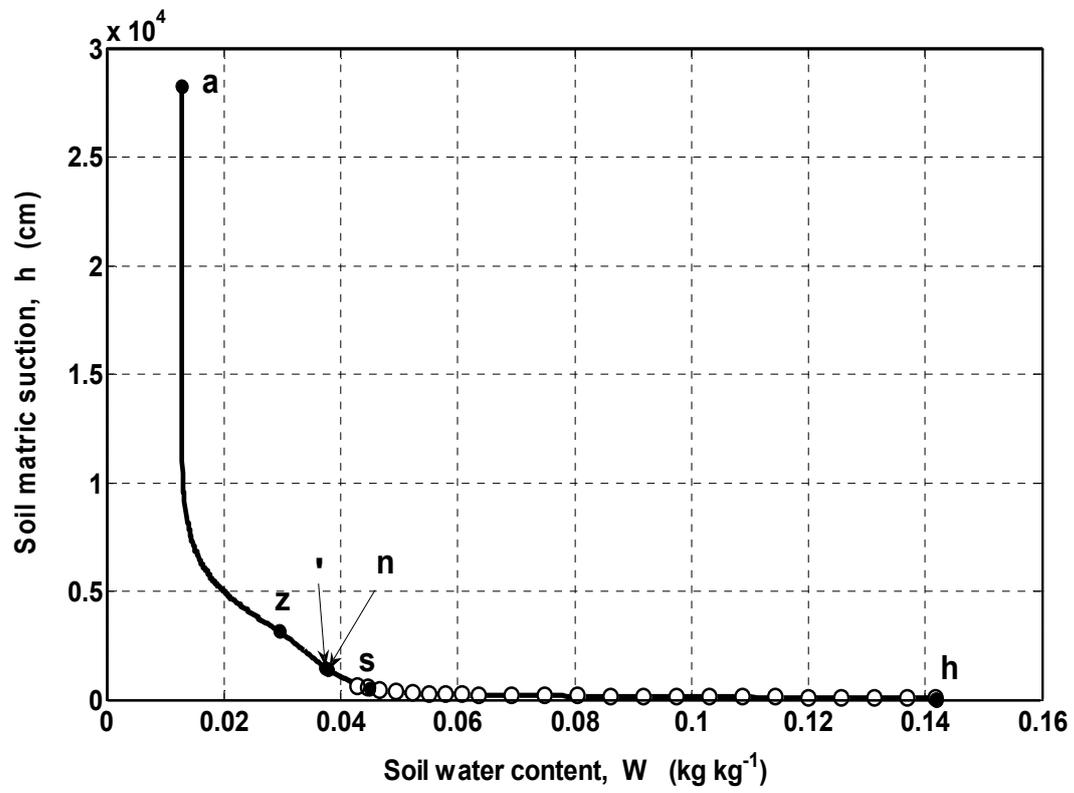

Fig.8a

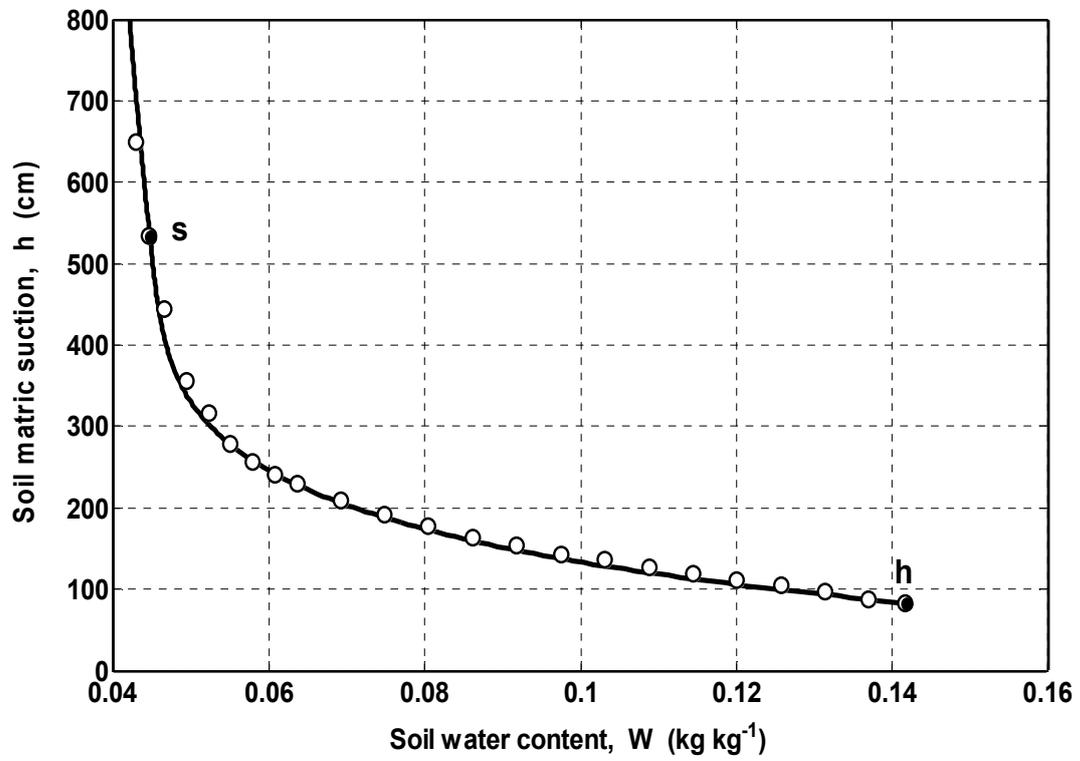

Fig.8b

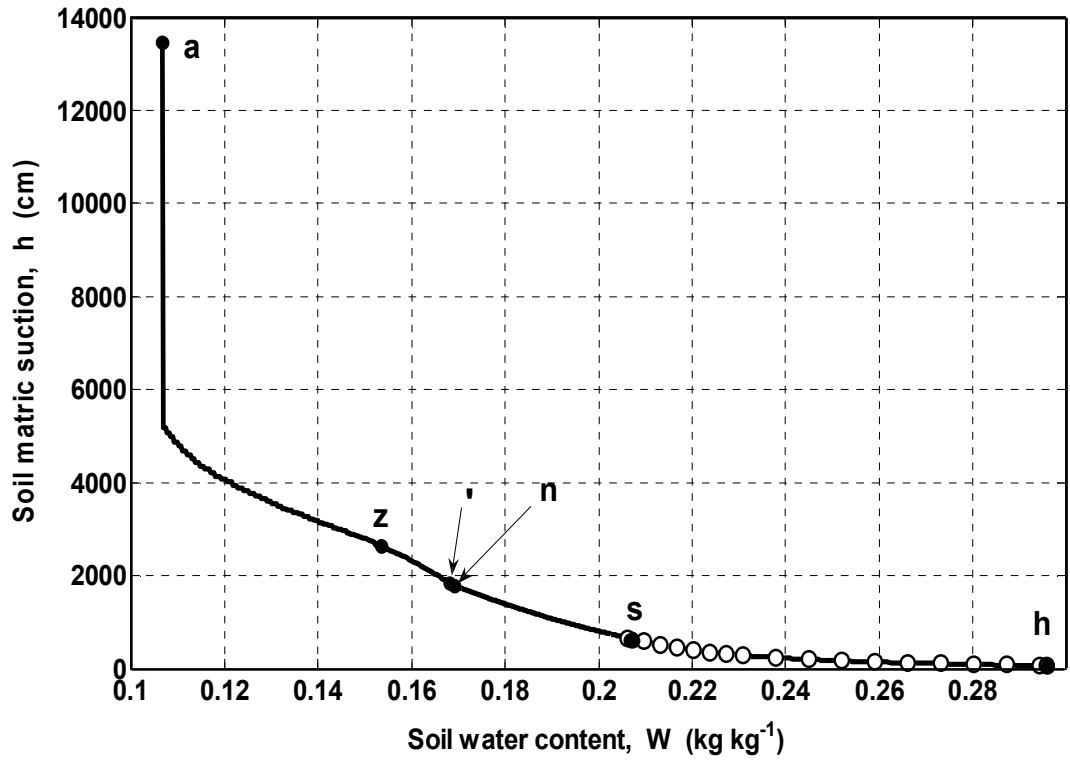

Fig.9a

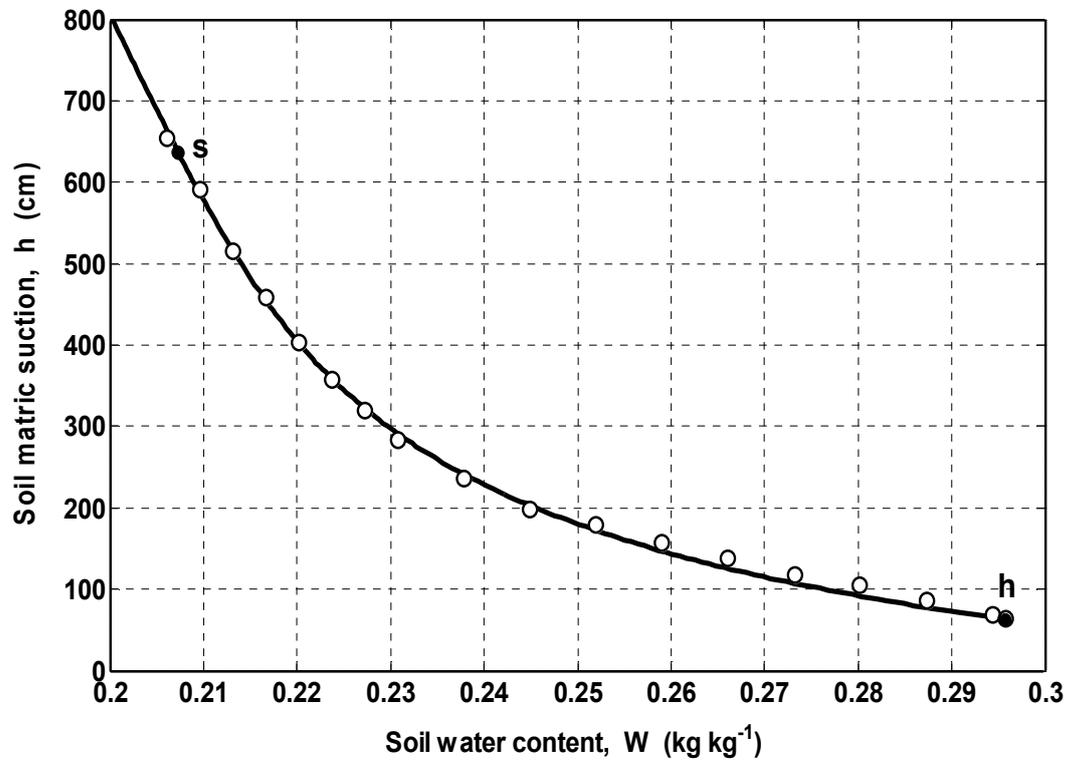

Fig.9b

**Table 1.** Input parameters[#] of the model [16,18] from which the observed shrinkage curves for soils from [12] are predicted. These parameters (i) contain three input parameters ($\rho_s$, $c$, $K$) for the soil water retention curve prediction, and (ii) determine (see [18]) three input parameters for this prediction, the $v_s$ and $v_z$ clay characteristics as well as the lacunar pore volume at maximum swelling, $U_{lph}=(W_h^*-W_h)/\rho_w$ (see Table 2)

| Soil | Data source | $Y_z$ | $W_h$ | $\rho_s$ | $c$ | $P_z$ | $K$ | $k$ | $W_h^*$ |
|---|---|---|---|---|---|---|---|---|---|
| | | dm³ kg⁻¹ | kg kg⁻¹ | kg dm⁻³ | | | | | kg kg⁻¹ |
| 1 | Cambisol from Fig.2a of [12] | 0.687 | 0.326 | 2.660 | 0.170 | 0 | 1.031 | 0.620 | 0.365 |
| 2 | Cambisol from Fig.2b of [12] | 0.674 | 0.285 | 2.660 | 0.140 | 0 | 1.035 | 0.829 | 0.321 |
| 3 | Vertisol from Fig.2d of [12] | 0.581 | 0.321 | 2.650 | 0.510 | 0.130 | 1.043 | 0.248 | 0.321 =$W_h$ |
| 4 | Fluvisol from Fig.2e of [12] | 0.558 | 0.142 | 2.660 | 0.090 | 0 | 2.709 | 0.875 | 0.184 |
| 5 | Fluvisol from Fig.2f of [12] | 0.598 | 0.296 | 2.650 | 0.420 | 0.021 | 1.190 | 0 | 0.296 =$W_h$ |
| 6 | Cambisol from Fig.5a of [12] | 0.720 | 0.296 | 2.660 | 0.150 | 0 | 1.046 | 0.815 | 0.362 |

[#]$Y_z$, oven-dried specific volume; $W_h$, maximum swelling water content; $\rho_s$ mean solid density; $c$, soil clay content; $P_z$, oven-dried structural porosity; $K$, aggregate/intra-aggregate mass ratio; $k$, lacunar factor; and $W_h^*$, water content that corresponds to filling in lacunar pores.

**Table 2.** Three input parameters[#] (together with other input parameters, $\rho_s$, $c$, $K$ from Table 1) from which the water retention curves for the six soils under consideration (see Figs.8a, b and 9a, b) as well as the estimated characteristic water contents of the soils[§] are predicted

| Soil | $v_s$ | $v_z$ | $U_{lph}$ | $W_a$ | $W_z$ | $W'$ | $W_n$ | $W_s$ |
|------|-------|-------|-----------|-------|-------|------|-------|-------|
|      |       |       | dm³kg⁻¹   | kg kg⁻¹ |     |      |       |       |
| 1 | 0.089 | 0.342 | 0.039 | 0.055 | 0.133 | 0.214 | 0.217 | 0.264 |
| 2 | 0.084 | 0.317 | 0.036 | 0.040 | 0.096 | 0.184 | 0.184 | 0.228 |
| 3 | 0.230 | 0.295 | 0     | 0.024 | 0.024 | 0.077 | 0.078 | 0.180 |
| 4 | 0.107 | 0.394 | 0.042 | 0.013 | 0.030 | 0.037 | 0.038 | 0.045 |
| 5 | 0.211 | 0.467 | 0     | 0.107 | 0.154 | 0.168 | 0.169 | 0.207 |
| 6 | 0.087 | 0.384 | 0.066 | 0.060 | 0.159 | 0.207 | 0.209 | 0.244 |

[#] $v_s$, relative volume of contributive clay solids; $v_z$, relative volume of contributive clay matrix in oven-dried state; and $U_{lph}$, specific volume of the clay lacunar pores within the intra-aggregate matrix at maximum swelling point.

[§] $W_a$, boundary of the exhausting capillary water; $W_z$, soil shrinkage limit; $W'$, "sewing" point; $W_n$, end point of basic shrinkage; and $W_s$, end point of soil structural shrinkage.

**Table 3. Estimated parameters[#] of the contributive clays for the soils under consideration that were used in the data analysis**

| Soil | $\zeta_z$ | $\zeta_n$ | $v_n$ | $Q_{nmin}$ | $Q_{nmax}$ | $P_n$ | $Q_n$ | $r^2$ | $\delta h$ |
|---|---|---|---|---|---|---|---|---|---|
|   |   |   |   |   |   |   |   |   | cm of $H_2O$ |
| 1 | 0.211 | 0.343 | 0.402 | 0.544 | 0.705 | 0.778 | 0.570 | 0.989 | 29.83 |
| 2 | 0.174 | 0.335 | 0.391 | 0.508 | 0.674 | 0.784 | 0.513 | 0.985 | 13.10 |
| 3 | 0.040 | 0.128 | 0.329 | 0.809 | 0.894 | 0.301 | 0.874 | 0.972 | 45.19 |
| 4 | 0.282 | 0.360 | 0.429 | 0.641 | 0.781 | 0.751 | 0.686 | 0.988 | 16.76 |
| 5 | 0.309 | 0.340 | 0.480 | 0.839 | 0.912 | 0.559 | 0.860 | 0.998 | 7.23 |
| 6 | 0.281 | 0.370 | 0.424 | 0.598 | 0.748 | 0.795 | 0.601 | 0.998 | 11.18 |

[#]$\zeta_z$, relative water content of clay at shrinkage limit; $\zeta_n$, relative water content of clay at air-entry point; $v_n$, relative clay volume at air-entry point; $Q_{nmin}$, possible minimum value of $Q_n$ for a clay; $Q_{nmax}$, possible maximum value of $Q_n$ for a clay; $P_n$, clay porosity at air-entry point; $Q_n$, clay $Q$ factor at air-entry point; $r^2$, goodness of fit for the best-fitted $Q_n$ value; and $\delta h$, estimate of standard error of experimental suction values

**Table 4. Estimates of all values[#] connected with boundary water content $\zeta_a$**

| Soil | $\zeta_a$ | $l_a$ | $\rho'_{min}(\zeta_a)$ | $\rho'_m(\zeta_a)$ | $\rho'_{min}(\zeta_a)/r_{mM}$ | $\rho'_m(\zeta_a)/r_{mM}$ | $F_a$ | $F_z$ | $L_a$ |
|------|-----------|-------|------------------------|---------------------|-------------------------------|----------------------------|-------|-------|-------|
|      |           |       | μm                     |                     |                               |                            |       |       | μm$^{-1}$ |
| 1    | 0.087     | 0.040 | 0.080                  | 1.446               | 0.028                         | 0.505                      | 0.313 | 0.762 | 7.830 |
| 2    | 0.074     | 0.036 | 0.073                  | 1.445               | 0.025                         | 0.492                      | 0.288 | 0.684 | 7.927 |
| 3    | 0.040     | 0.075 | 0.190                  | 1.337               | 0.063                         | 0.445                      | 0.472 | 0.472 | 6.260 |
| 4    | 0.122     | 0.051 | 0.102                  | 1.444               | 0.037                         | 0.529                      | 0.381 | 0.873 | 7.506 |
| 5    | 0.214     | 0.107 | 0.214                  | 1.406               | 0.083                         | 0.546                      | 0.660 | 0.952 | 6.179 |
| 6    | 0.106     | 0.041 | 0.082                  | 1.453               | 0.030                         | 0.528                      | 0.324 | 0.864 | 7.898 |

[#]$\zeta_a$, relative clay water content corresponding to maximum adsorbed film; $l_a$, maximum thickness of adsorbed water film; $\rho'_{min}(\zeta_a)$, minimum $\rho'$ value in clay matrix at $\zeta=\zeta_a$; $\rho'_m(\zeta_a)$, maximum $\rho'$ value in clay matrix at $\zeta=\zeta_a$; $\rho'_{min}(\zeta_a)/r_{mM}$, minimum relative internal size of pore-tube cross-sections at $\zeta=\zeta_a$; $\rho'_m(\zeta_a)/r_{mM}$, maximum relative internal size of pore-tube cross-sections at $\zeta=\zeta_a$; $F_a$, clay saturation degree at $\zeta=\zeta_a$; $F_z$, clay saturation degree at shrinkage limit; and $L_a$, summary perimeter of pore tubes (per unit surface area of their cross-section) containing only the maximum adsorbed water film at $\zeta=\zeta_a$.